\documentclass[12pt,a4paper]{article}

\usepackage[colorlinks]{hyperref}
\usepackage{amsmath}
\usepackage{amsthm}
\usepackage{amsfonts}
\usepackage{amssymb}
\usepackage{mathabx}
\usepackage{bussproofs}
\usepackage{lineno}
\usepackage{ cmll } 
\usepackage{graphicx}
\usepackage{relsize}

\usepackage{url}
\usepackage{enumitem}
\usepackage{xcolor}
\usepackage{mathrsfs}
\usepackage{colortbl}
\usepackage{colonequals}
\usepackage[all]{xy}
\usepackage{tikz}
\usetikzlibrary{positioning,arrows,calc}
\usepackage{graphicx}

\theoremstyle{plain} 
\newtheorem{thm}{Theorem}[section]
\newtheorem{lem}[thm]{Lemma}

\newtheorem{cor}[thm]{Corollary}
\theoremstyle{definition}
\newtheorem{dfn}[thm]{Definition}
\newtheorem{exam}[thm]{Example}
\newtheorem{rem}[thm]{Remark}

\usetikzlibrary{positioning,arrows,calc}
\usepackage[ruled,lined,linesnumbered]{algorithm2e}
\SetKwComment{Comment}{/* }{ */}

\usepackage{titlesec}

\titleformat*{\section}{\normalfont\fontfamily{phv}\fontsize{16}{19}\bfseries}
\titleformat*{\subsection}{\normalfont\fontfamily{put}\fontsize{12}{15}\bfseries}
\titleformat*{\subsubsection}{\normalfont\fontfamily{put}\fontsize{14}{17}\selectfont}

\usepackage{booktabs,array}
\newcolumntype{P}[1]{>{\raggedright\arraybackslash}p{\dimexpr#1\linewidth-2\tabcolsep}}

\def\FL_e{\mathrm{FL_e}}

\def\CPC{\mathrm{CPC}}

\def\4D{\mathsf{4D}}

\def\S4{\mathsf{S4}}
\def\IPC{\mathsf{IPC}}
\def\CPC{\mathsf{CPC}}
\def\ruleK{\mathsf{K(4)}_\alpha^{J,L}}
\def\ruleD{\mathsf{D(4)}_\alpha^{J, L}}
\def\phi{\varphi}
\def\IPC{\mathsf{IPC}}
\def\CPC{\mathsf{CPC}}

\newcommand{\calS}{\mathcal{S}}
\newcommand{\tGamma}{\tilde{\Gamma}}
\newcommand{\tSigma}{\tilde{\Sigma}}
\newcommand{\tPi}{\tilde{\Pi}}
\newcommand{\tDelta}{\tilde{\Delta}}
\newcommand{\tLambda}{\tilde{\Lambda}}
\def\phi{\varphi}

\begin{document}

\title{Universal Proof Theory: Semi-analytic Rules and Craig Interpolation} 
\author{Amirhossein Akbar Tabatabai$^{\tiny{a}}$\footnote{amir.akbar@gmail.com. Supported by the GA\v{C}R grant 23-04825S, the FWF project P 33548 and the Czech Academy of Sciences (RVO 67985840).}, Raheleh Jalali$^{\tiny{b}}$\footnote{rahele.jalali@gmail.com. Supported by the Czech Science
Foundation project 22-01137S, RVO: 67985840, and the grant 639.073.807 of the Netherlands Organisation for Scientific Research.} \\ \small{$^{\tiny{a}}$Institute of Mathematics, Czech Academy of Sciences}\\ \small{$^{\tiny{b}}$Institute of Computer Science, Czech Academy of Sciences}
}
\maketitle 

\begin{abstract}
We provide a general and syntactically defined family of sequent calculi, called \emph{semi-analytic}, to formalize the informal notion of a ``nice" sequent calculus. We show that any sufficiently strong (multimodal) substructural logic with a semi-analytic sequent calculus enjoys the Craig Interpolation Property, CIP. As a positive application, our theorem provides a uniform and modular method to prove the CIP for several multimodal substructural logics, including many fragments and variants of linear logic. More interestingly, on the negative side, it employs the lack of the CIP in almost all substructural, superintuitionistic and modal logics to provide a formal proof for the well-known intuition that almost all logics do not have a ``nice" sequent calculus. More precisely, we show that many substructural logics including $\mathsf{UL^-}$, $\mathsf{MTL}$,  $\mathsf{R}$, \textsf{Ł}$_n$ (for $n \geq 3$),  \textsf{G}$_n$ (for $n \geq 4$),  and 
almost all extensions of $\mathsf{IMTL}$, $ \textsf{Ł}$, $\mathsf{BL}$, $\mathsf{RM^e}$, $\mathsf{IPC}$, $\mathsf{S4}$, and $\mathsf{Grz}$  (except for at most 1, 1, 3, 8, 7, 37, and 6 of them, respectively) do not have a semi-analytic calculus.


\end{abstract}
\section{Introduction}
Proof systems have the main role in all proof-theoretic investigations, from Gentzen's consistency proof and Kreisel's proof mining program to the characterization of the admissible rules of a logical system and the establishment of its decidability. To investigate such logical properties, proof theorists usually design \emph{specific} proof systems for their specific purposes. For them, a proof system is just a technical tool to establish the intended logical properties of a given mathematical theory and compared to these logical properties, proofs and proof systems are only second-class citizens, far from the independent mathematical objects that they deserve to be. Fortunately, in recent years an approach has emerged in proof theory that intends to shift the focus from the classical goal of design-and-mine the \emph{specific} proof systems to the more modern investigation of the form and properties of a \emph{generic} proof system  \cite{Cia,Iemhoff1,Iemhoff}.
We call this emerging program \emph{universal proof theory}\footnote{We are grateful to Masoud Memarzadeh for this terminological suggestion.}, a name we hope to be reminiscent of \emph{universal algebra}, a theory studying the generic behaviour of algebraic structures. 

The first step in this program is formalizing what \emph{``nice"} proof systems are and for what logics they \emph{exist}. The hope is to formalize the well-known intuition that nice proof systems are rare objects to find and hence in almost all possible cases, they are simply non-existent. To prove such negative results, after fixing a \emph{family} of logics and
providing a candidate for the nice proof systems, a natural strategy is finding an \emph{invariant property} for the logics in the family such that if a logic has a nice proof system, it must have that property. If this property is also rare, one can finally prove that nice proof systems are also rare. The idea of employing such a strategy to prove the non-existence of proof systems of a certain general form was first proposed by Iemhoff \cite{Iemhoff1,Iemhoff}. The family she considered is the class of (modal) superintuitionistic logics, her invariant property is the uniform interpolation property (UIP), and her nice sequent calculi are the calculi consisting of the usual axioms of the sequent calculus $\mathbf{LJ}$, for the intuitionistic propositional logic, and a special form of rules she calls \emph{focused}. Roughly speaking, a rule is focused when it has only one main formula in the conclusion and satisfies:

$\bullet$ 
\emph{focused condition}: both the main and active formulas occur on the same side of the sequent, i.e., either in the antecedent or the succedent of the sequents in the rule, and

$\bullet$ 
\emph{occurrence-preservation:}
any atom occurring in any active formula also occurs in the main formula.

\noindent For instance, the usual conjunction and disjunction rules are focused while the implication rules are not, as they violate the focused condition. Iemhoff proves that if a (modal) superintuitionistic logic has a terminating sequent calculus of this form, then the logic enjoys the UIP. To establish the intended non-existence result, as the UIP is a rare property for a logic in the chosen family, she concludes that all superintuitionistic logics, except at most seven of them and the logics $\mathsf{K4}$ and $\mathsf{S4}$ cannot have such a terminating calculus.

In this paper, we also focus on the non-existence of nice proof systems. In this regard, we substantially generalize Iemhoff's setting as described below. First, we generalize the focused axioms of \cite{Iemhoff} to a very general form. We also generalize the
focused rules to what we call \emph{semi-analytic}, by relaxing the focused condition and generalizing the form to exhaust almost all combinatorial possibilities. Therefore, semi-analytic rules cover many rules in the literature including the focused and the implication rules together with any combination of context-sharing and context-combining attitudes in substructural logic. We also consider a multimodal version of the usual modal rules to address the modalities known as the subexponentials in the linear logic community \cite{danos,nigam,nigam2011,olarte}. Second, for our family of logics, we lower the base logic from the intuitionistic logic to the substructural logics $\mathsf{FL_e}$ (Full Lambek logic with the exchange rule) and $\mathsf{IMALL}$ (Intuitionistic Multiplicative-Additive Linear Logic) to extend the applicability of our final result to substructural and linear logics, as well. Finally, for the invariant property, we consider the Craig Interpolation Property (CIP) that weakens the invariant property and hence strengthens the negative result. The main theorem we prove is that
any (multimodal) logic extending the multiplicative-additive linear logic $\mathsf{MALL}$ (resp. $\mathsf{IMALL}$) that has a multi-conclusion (resp. single-conclusion) sequent calculus only consisting of multi-conclusion (resp. single-conclusion) semi-analytic rules, focused axioms and basic multimodal rules has the CIP. A similar result is also proved replacing $\mathsf{MALL}$ (resp. $\mathsf{IMALL}$)  with $\mathsf{CFL_e}$ (resp. $\mathsf{FL_e}$) as the base logic. For the definitions of these logics, see \cite{ono}.

As a positive application, our theorem provides a uniform method to establish the CIP for several (multimodal) substructural logics. The method is based on checking the form of the axioms and rules of the sequent calculus and as it is a local and syntactical procedure, the technique is quite modular and user-friendly. Using this method, we reprove the CIP for many logics, including intuitionistic propositional logic, classical propositional logic, substructural logics
$\mathsf{FL_e}$, $\mathsf{FL_{ew}}$, 
$\mathsf{FL_{ec}}$, their classical versions $\mathsf{CFL_e}$, $\mathsf{CFL_{ew}}$, $\mathsf{CFL_{ec}}$, the bounded versions $\mathsf{IMALL}$, $\mathsf{MALL}$ and several of their modal extensions including linear logics $\mathsf{ILL}$, $\mathsf{CLL}$, $\mathsf{ELL}$, and their affine and relevant versions. Our main new result though is the CIP for the general setting of multimodal substructural logics introduced in \cite{elaine} that combines basic substructural logics with multimodalities, each of which can be of its own almost independent type, chosen from the set of the types $\{\mathrm{K}, \mathrm{KD}, \mathrm{KT}, \mathrm{K4}, \mathrm{KD4}, \mathrm{S4}\}$. 

On the negative side, we reach our main goal. We show that any logic extending $\mathsf{MALL}$ (resp. $\mathsf{IMALL}$) and lacking the CIP cannot have a multi-conclusion (resp. single-conclusion) sequent calculus consisting only of semi-analytic rules, focused axioms and some basic multimodal rules. A similar result is also in place replacing $\mathsf{MALL}$ (resp. $\mathsf{IMALL}$) with $\mathsf{CFL_e}$ (resp. $\mathsf{FL_e}$).
The logics that have no sequent calculus of the described form include many substructural, relevant and semilinear logics such as $\mathsf{UL^-}$, $\mathsf{MTL}$,  $\mathsf{R}$, \textsf{Ł}$_n$ (for $n \geq 3$),  \textsf{G}$_n$ (for $n \geq 4$),  and 
almost all extensions of $\mathsf{IMTL}$, $ \textsf{Ł}$, $\mathsf{BL}$, $\mathsf{RM^e}$, $\mathsf{IPC}$, $\mathsf{S4}$, and $\mathsf{Grz}$,  (except for at most 1, 1, 3, 8, 7, 37, and 6 of them, respectively).

\section{Preliminaries}

In this section, we recall some basics from the proof theory of (modal) substructural and linear logics. For more, the reader may consult \cite{ono,tro92,troelstra,chagrov}. We work with two families of languages; the \emph{bounded} language  $\mathcal{L}_I^b=\{\wedge, \vee, \to, *, 0, 1, \top, \bot\} \cup \{!^{\alpha}\mid \alpha \in I\}$ and the \emph{unbounded} language $\mathcal{L}_I^{u}=\mathcal{L}_I^b \setminus \{\top, \bot\}$, where $*$ is called \emph{fusion} and $I$ is a (possibly empty) set of indices and $!^{\alpha}$ is a modality, for any $\alpha \in I$. To refer to either of these languages, we use the variable $\mathcal{L} \in \{\mathcal{L}_I^b, \mathcal{L}_I^{u}\}$.  We define \emph{$\mathcal{L}_I^b$-formulas} by the following grammar:
\begin{center}
$F::= p \mid 0 \mid 1 \mid \bot \mid \top \mid F_1 \wedge F_2 \mid F_1 \vee F_2 \mid F_1 \to F_2 \mid F_1 * F_2 \mid !^\alpha F$
\end{center}
and $\mathcal{L}_I^{u}$-formulas are defined similarly. When the language $\mathcal{L}$ is clear from the context, we call an $\mathcal{L}$-formula simply a formula. Multisets of formulas are defined as usual. We also use \emph{multiset variables} (also called \emph{contexts}) to talk about a generic multiset, as it is common in proof theory. Since we cover some variants of linear logic, it is worth recalling the translation between our language and the one used by Girard \cite{Girard}, see Table \ref{table: notation}. 
We reserve the letters $\alpha, \beta, \gamma$ (possibly with indices) to denote different modalities. The \emph{atomic formulas (atoms)} are denoted by $p, q$ and they are sometimes called the \emph{variables}. We denote $\mathcal{L}$-formulas by  $\phi, \psi, \theta, \mu, \nu, \rho$ (possibly with indices). We use the bar notation as in $\overline{\phi}, \overline{\psi}, \ldots$, the curly brackets as in $\{\phi_1, \dots, \phi_n\}$, and the capital Greek letters $\Gamma, \Delta, \ldots$ to denote finite (possibly empty) multisets of $\mathcal{L}$-formulas. The capital Greek letters with a tilde as in $\tGamma, \tDelta, \ldots$ are reserved for the multiset variables. 
Denote the number of elements of the multiset $\Gamma$ by $|\Gamma|$. By $\Gamma, \Delta$ we mean $\Gamma \cup \Delta$. 
Define $\neg \phi:=\phi \to 0$, $\phi+\psi:=\neg (\neg \phi * \neg \psi)$, $\neg \Gamma=\{\neg \phi \mid \phi \in \Gamma\}$, $!^\alpha \Gamma=\{!^\alpha \phi \mid \phi \in \Gamma\}$, and $!^\alpha \emptyset:=\emptyset$. If $\Gamma=\{\phi_1, \dots, \phi_n\}$, define $\bigast \Gamma: =\phi_1 * \dots * \phi_n$, $\bigplus \Gamma: =\phi_1 + \dots + \phi_n$ and set $\bigast \emptyset := 1$ and $\bigplus \emptyset := 0$. Define $\phi^n$ recursively by $\phi^0=1$ and $\phi^{n+1}=\phi^{n}*\phi$. 
The \emph{set of the variables} of a formula $\phi$, denoted by $V(\phi)$, is defined recursively by $V(c)=\emptyset$, for any constant $c \in \{\top, \bot, 0, 1\}$; $V(p)=p$ for any atom $p$; $V(\phi \circ \psi)=V(\phi) \cup V(\psi)$, for any  $\circ \in \{\wedge, \vee, \to, * \}$; and  $V(!^\alpha \phi)= V(\phi)$, for any $\alpha \in I$. Define $V(\Gamma)= \bigcup_{\phi \in \Gamma} V(\phi) $, for a multiset $\Gamma$.
For $\mathcal{L}=\mathcal{L}^b_I$, a \emph{substitution} $\sigma$ on $\mathcal{L}$ is a function  
mapping $\mathcal{L}$-formulas to $\mathcal{L}$-formulas such that:
$\sigma(c)=c$, $\sigma(\phi \circ \psi)=\sigma(\phi) \circ \sigma(\psi)$, and $\sigma(!^\alpha \phi)=!^\alpha \sigma(\phi)$, for any $\mathcal{L}$-formulas $\phi$ and $\psi$, and any $c \in \{0, 1, \top, \bot\}$,  $\circ \in \{\wedge, \vee, \to, *\}$ and $\alpha \in I$. For $\mathcal{L}=\mathcal{L}^u_I$, the definition is similar, replacing the set $\{0, 1, \top, \bot \}$ by $\{0, 1\}$. Note that $\sigma (\phi(p_1,\dots,p_n))=\phi(\sigma(p_1),\dots,\sigma(p_n))$, for any formula $\phi$, where $p_i$'s are all the atoms 
occurring in $\phi$. For a multiset $\Gamma$ and a substitution $\sigma$, define $\sigma(\Gamma)$ as $\{\sigma(\phi) \mid \phi \in \Gamma\}$. 
\begin{table}[t!]
\begin{center}
\scalebox{0.9}{{\begin{tabular}{ |c| c| c| c| }
\hline
linear logic & our notation \\
\hline
$\multimap$ &  $\to$ \\
\hline
$\&$ &  $\wedge$ \\
\hline
$\oplus$ &  $\vee$ \\
\hline
$\otimes$ &  $*$ \\
\hline
$\parr$ &  $+$ \\
\hline
$(-)^\bot$ &  $\neg$ \\
\hline
$\bot$ &  $0$ \\
\hline
$1$ &  $1$ \\
\hline
$\top$ &  $\top$ \\
\hline
$0$ &  $\bot$ \\
\hline
\end{tabular}}}
\caption{{\small Translation between linear logic and our notation.}}\label{table: notation}
\end{center}
\end{table}

\begin{dfn} \label{DfnLogic} 
A \emph{logic} $L$ over $\mathcal{L}$ is a set of $\mathcal{L}$-formulas closed under substitution, modus ponens ($\phi, \phi \to \psi \in L$ implies $\psi \in L$), and adjunction ($\phi, \psi \in L$ implies $\phi \wedge \psi \in L$). We denote the language of the logic $L$ by $\mathcal{L}_L$. 
For logics $L$ and $L'$ such that $\mathcal{L}_L \subseteq \mathcal{L}_{L'}$, we say that $L'$ is an \emph{extension} of $L$ (or $L'$ \emph{extends} $L$) if $L \subseteq L'$.
\end{dfn}

A \emph{multi-conclusion sequent} (\emph{sequent}, for short) over $\mathcal{L}$, denoted by the letter $S$ (possibly with indices), is an expression in the form 
$\Gamma \Rightarrow \Delta$, where $\Gamma$ and $\Delta$ are finite multisets of $\mathcal{L}$-formulas. It is \emph{single-conclusion} when $|\Delta| \leq 1$. For the sequent $S=(\Gamma \Rightarrow \Delta)$ over $\mathcal{L}$, the \emph{interpretation} of $S$, denoted by $I(S)$, is defined as the $\mathcal{L}$-formula $\bigast \Gamma \to \bigplus \Delta$ and the \emph{antecedent} (resp. \emph{succedent}) of $S$, denoted by $S^a$ (resp. $S^s$), is defined as $\Gamma$ (resp. $\Delta$). 
By $\Gamma \Leftrightarrow \Delta$, we mean $\Gamma \Rightarrow \Delta$ and $\Delta \Rightarrow \Gamma$.
\emph{Multi-conclusion meta-sequents} (\emph{meta-sequents}, for short) over $\mathcal{L}$, denoted by the calligraphic letter $\calS$ (possibly with indices), are defined similar to sequents, except that for a multiset variable $\tSigma$, we also allow $\tSigma$ and $!^\alpha \tSigma$ in the antecedent and/or succedent, for each $\alpha \in I$. 
Therefore, any meta-sequent over $\mathcal{L}$ is of the form
\begin{center}
    $!^{\alpha_1}\tGamma_1, \ldots, !^{\alpha_m}\tGamma_m, \tPi_{1}, \ldots, \tPi_n, \overline{\mu} \Rightarrow \bar{\nu}, !^{\beta_1}\tDelta_1, \ldots, !^{\beta_r}\tDelta_r, \tLambda_1, \ldots, \tLambda_s \quad (1)$
\end{center}
where the $\mathcal{L}$-formulas in $\bar{\mu}$ and $\bar{\nu}$ are called the \emph{formulas of the meta-sequent}. $\calS^{a}$ and $\calS^{s}$ are defined similarly.   
A \emph{substitution} for a sequent or meta-sequent, denoted by $\sigma(S)$ and $\sigma(\calS)$, respectively, is defined in the usual way, mapping $\mathcal{L}$-formulas to $\mathcal{L}$-formulas and multiset variables to multisets of $\mathcal{L}$-formulas. A meta-sequent $\calS$ of the form specified in $(1)$ is called \emph{single-conclusion} if  $\calS^s=\bar{\nu}$, where $|\bar{\nu}| \leq 1$ or $\calS^s=!^{\beta_1}\tDelta_1, \ldots, !^{\beta_r}\tDelta_r, \tLambda_1, \ldots, \tLambda_s$.  The intuition behind a single-conclusion meta-sequent is that after the substitution, the succedent has at most one formula. Especially in the latter case, we are allowed to substitute at most one formula in one of the contexts and all the others must be substituted with the empty multiset. 
Such a setting is required to define the multiplicative rules, like the left implication or right fusion rules, where the contexts of the premises are merged in the conclusion.

\noindent Define the \emph{single-conclusion version} of a meta-sequent $\calS$ of the form $(1)$ as:
\begin{itemize}
\item 
if $|\bar{\nu}|>1$ then $\calS$ has no single-conclusion version;
\item 
if $\bar{\nu}=\nu$, then its single-conclusion version is:
\begin{center}
   $!^{\alpha_1}\tGamma_1, \ldots, !^{\alpha_m}\tGamma_m, \tPi_{1}, \ldots, \tPi_n, \overline{\mu} \Rightarrow \nu$;
\end{center}
\item 
if $\bar{\nu}=\emptyset$, then its single-conclusion version is $\calS$ itself.
\end{itemize}
A \emph{multi-conclusion rule} (\emph{rule}, for short) over $\mathcal{L}$ is an expression
\begin{center} 
 \AxiomC{${\calS}_1, \dots, {\calS}_n$}
 \UnaryInfC{${\calS}$}
 \DisplayProof \quad (2)
\end{center} 
where ${\calS}, {\calS}_1, \ldots, {\calS}_n$ are meta-sequents over $\mathcal{L}$. Meta-sequents above the line are the \emph{premises} and the one below is the \emph{conclusion} of the rule. An \emph{axiom} is a rule with no premises.
The formulas in the conclusion are called the \emph{main} formulas and the formulas in the premises are called the \emph{active} formulas of the rule. Usually, we work with the rules with exactly one main formula. A rule is called \emph{single-conclusion} if all of its meta-sequents are single-conclusion. Define the \emph{single-conclusion version} of a rule $\mathcal{R}$ of the form $(2)$ as follows:
\begin{itemize}
\item[$(i)$] 
if $\calS$ or any of the $\calS_i$'s for $1\leq i \leq n$ does not have a single-conclusion version, then $\mathcal{R}$ does not have a single-conclusion version; otherwise,
\item[$(ii)$]  
replace each meta-sequent in $\mathcal{R}$ by its single-conclusion version. Then, if there is a context in the conclusion that does not appear in any of the premises, delete it. The result is the single-conclusion version of $\mathcal{R}$.
\end{itemize}
For instance, the single-conclusion version of the rule $(L \to)$ is $(L \to)'$:
\begin{center}
\begin{tabular}{c c}
\AxiomC{$\tGamma \Rightarrow p, \tDelta$}
 \AxiomC{$\tSigma, q \Rightarrow \tLambda$}
 \RightLabel{\scriptsize{$(L \to)$}}
 \BinaryInfC{$\tGamma, \tSigma, p \to q \Rightarrow \tDelta, \tLambda$}
 \DisplayProof \quad \quad
 &
\AxiomC{$\tilde{\Gamma} \Rightarrow p$}
 \AxiomC{$\tilde{\Sigma}, q \Rightarrow \tilde{\Lambda}$}
 \RightLabel{\scriptsize{$(L \to)'$}}
 \BinaryInfC{$\tilde{\Gamma}, \tilde{\Sigma}, p \to q \Rightarrow \tilde{\Lambda}$}
 \DisplayProof
\end{tabular}
\end{center}
while the rule
\begin{center}
\AxiomC{$\tGamma \Rightarrow p, p, \tDelta$}
 \UnaryInfC{$\tGamma, \tSigma \Rightarrow p, \tDelta$}
 \DisplayProof
\end{center}
does not have a single-conclusion version.

An \emph{instance} of a rule of the form $(2)$ is defined as $\sigma(\calS_1), \dots, \sigma(\calS_n)/ \sigma(\calS)$, where $\sigma$ is a substitution function on $\mathcal{L}$.
For a single-conclusion rule, the instance is called single-conclusion if in addition, each $\sigma(\calS_1), \dots, \sigma(\calS_n),$ $\sigma(\calS)$ is single-conclusion.
By a \emph{multi-conclusion sequent calculus} (\emph{calculus}, for short) $G$ over $\mathcal{L}$, we mean a set of rules over $\mathcal{L}$.  We denote the language of the calculus $G$ by $\mathcal{L}_G$. A calculus is \emph{single-conclusion} if all of its rules are single-conclusion. The \emph{single-conclusion version} of a calculus $G$ is defined as the set of all single-conclusion versions of the rules of $G$, if they exist, i.e., not including the rules which do not have a single-conclusion version.
A (single-conclusion) sequent $S$ is \emph{provable} from a set $\mathcal{C}$ of (single-conclusion) sequents in a (single-conclusion) calculus $G$, denoted by $\mathcal{C} \vdash_G S$, if there is a finite tree with (single-conclusion) sequents as the labels of its nodes such that the label of the root is $S$, labels of the leaves are (single-conclusion) instances of the axioms of $G$ or the elements of $\mathcal{C}$,
and in each node the set of the labels of the children of the node together with the label of the node itself, constitute a (single-conclusion) instance of a rule in $G$. This finite tree is called the \emph{proof} of $S$ from $\mathcal{C}$ in $G$. We denote $\varnothing \vdash_G S$ by $G \vdash S$ and we say $S$ is provable in $G$. Formulas $\phi$ and $\psi$ are \emph{provably equivalent} in a calculus, when $\phi \Leftrightarrow \psi$ is provable in it.
We denote specific calculi, such as $\mathbf{FL_e}$, by bold-faced letters and specific logics, such as $\mathsf{FL_e}$, by sans-serif letters. 

\begin{dfn} \label{CalculusExtension}
Let $G$ and $H$ be two (single-conclusion) sequent calculi such that $\mathcal{L}_G \subseteq \mathcal{L}_{H}$. We say that $H$ is an \emph{extension} of $G$ if every (single-conclusion) instance of the axioms of $G$ is provable in $H$ and every rule of $G$ is \emph{admissible} in $H$, i.e., for any (single-conclusion) instance of a rule of $G$, if the premises are provable in $H$ then so is its conclusion.
\end{dfn}

\begin{dfn} \label{CalculusForLogic}
Let $L$ be a logic and $G$ be a (single-conclusion) calculus over $\mathcal{L} \in \{\mathcal{L}_I^b, \mathcal{L}_I^{u}\}$. We say $G$ is a \emph{sequent calculus for $L$} or $L$ is \emph{the logic of} $G$, when
$G \vdash \Gamma \Rightarrow \Delta$
if{f}
$(\bigast \Gamma \rightarrow \bigplus \Delta) \in L$, for any (single-conclusion) sequent $\Gamma \Rightarrow \Delta$ over $\mathcal{L}$.
\end{dfn}
Consequently, if $G$ is a calculus for $L$, we have 
$G \vdash \phi \Rightarrow \psi \text{ iff }  (\phi \to \psi) \in L$, for any formulas $\phi$ and $\psi$.  
\subsection{Substructural and linear  calculi}\label{subsec: 2.1}
We recall some well-known calculi for substructural and linear logics, setting $I=\varnothing$ or $I=\{\alpha\}$. For the empty $I$, define $\mathbf{CFL_e}$ over $\mathcal{L}_{\varnothing}^u$ as the calculus consisting of all the axioms and rules in Table \ref{tableFLe}. 
\begin{table}[t]
  \centering
\resizebox{\columnwidth}{!}{
 \begin{tabular}{ p{3cm} p{3cm} p{3cm} p{3cm} }
   \multicolumn{2}{c}{ \small  $p \Rightarrow p \;\;$ \scriptsize$(id)$}  & \hspace{-3em} \small $\Rightarrow 1 \;\;$ & \hspace{-3em} \small $0 \Rightarrow \;\;$
    \vspace{7pt}\\
\multicolumn{2}{c}{\hspace{5em}\small
\AxiomC{$\tGamma \Rightarrow \tDelta$}
 \RightLabel{\scriptsize{$(1 w)$}} 
 \UnaryInfC{$\tGamma, 1 \Rightarrow \tDelta$}
 \DisplayProof}
 & 
 \multicolumn{2}{c}{\hspace{-5em}\small \AxiomC{$\tGamma \Rightarrow \tDelta$}
 \RightLabel{\scriptsize{$(0 w)$}} 
 \UnaryInfC{$\tGamma \Rightarrow 0 , \tDelta$}
 \DisplayProof}
  \vspace{7pt}\\
 \small  
 \AxiomC{$\tGamma, p \Rightarrow \tDelta$}
 \RightLabel{\scriptsize{$(L \wedge_1)$} }
 \UnaryInfC{$\tGamma, p \wedge q \Rightarrow \tDelta$}
 \DisplayProof
&
\small
\AxiomC{$\tGamma, q \Rightarrow \tDelta$}
 \RightLabel{\scriptsize{$(L \wedge_2)$} }
 \UnaryInfC{$\tGamma, p \wedge q \Rightarrow \tDelta$}
 \hspace*{2em}\DisplayProof
 & 
\multicolumn{2}{c}{ \small
\AxiomC{$\tGamma \Rightarrow p, \tDelta$}
 \AxiomC{$\tGamma \Rightarrow q, \tDelta$}
 \RightLabel{\scriptsize{$(R \wedge)$} }
 \BinaryInfC{$\tGamma \Rightarrow p \wedge q, \tDelta$}
  \hspace*{2em}\DisplayProof}
   \vspace{7pt}
   \\
 \small  
  \AxiomC{$\tGamma, p \Rightarrow \tDelta$}
 \AxiomC{$\tGamma, q \Rightarrow \tDelta$}
 \RightLabel{\scriptsize{$(L \vee)$} }
 \BinaryInfC{$\tGamma, p \vee q \Rightarrow \tDelta$}
 \DisplayProof
& 
\multicolumn{2}{c}{\small 
 \AxiomC{$\tGamma \Rightarrow p, \tDelta$}
 \RightLabel{\scriptsize{$(R \vee_1)$} }
 \UnaryInfC{$\tGamma \Rightarrow p \vee q, \tDelta$}
 \hspace*{2em}\DisplayProof}
 & 
 \small  \AxiomC{$\tGamma \Rightarrow q, \tDelta$}
 \RightLabel{\scriptsize{$(R \vee_2)$} }
 \UnaryInfC{$\tGamma \Rightarrow p \vee q, \tDelta$}
 \hspace*{-3em}\DisplayProof
  \vspace{7pt} \\
    \multicolumn{2}{c}{
\small   \AxiomC{$\tGamma, p, q \Rightarrow \tDelta$}
 \RightLabel{\scriptsize{$(L*)$}}
 \UnaryInfC{$\tGamma, p * q \Rightarrow \tDelta$}
 \hspace*{2em}\DisplayProof}
&
\multicolumn{2}{c}{
\small  
 \AxiomC{$\tGamma \Rightarrow p, \tDelta$}
 \AxiomC{$\tSigma \Rightarrow q, \tLambda$}
 \RightLabel{\scriptsize{$(R*)$} }
 \BinaryInfC{$\tGamma, \tSigma \Rightarrow p * q, \tDelta, \tLambda$}
 \hspace*{-1em}\DisplayProof}
 \vspace{7pt}\\
    \multicolumn{2}{c}{
\small   \AxiomC{$\tGamma \Rightarrow p, \tDelta$}
 \AxiomC{$\tSigma, q \Rightarrow \tLambda$}
 \RightLabel{\scriptsize{$(L \to)$} }
 \BinaryInfC{$\tGamma, \tSigma, p \to q \Rightarrow \tDelta, \tLambda$}
 \hspace*{2em}\DisplayProof}
&
\multicolumn{2}{c}{
\small  \AxiomC{$\tGamma, p \Rightarrow q, \tDelta$}
 \RightLabel{\scriptsize{$(R \to)$} }
 \UnaryInfC{$\tGamma \Rightarrow p \to q, \tDelta$}
 \hspace*{-1em}\DisplayProof}
 \end{tabular} }\\
 \vspace{7pt}
 \AxiomC{$\tGamma \Rightarrow p, \tDelta$}
 \AxiomC{$\tSigma, p \Rightarrow \tLambda$}
 \RightLabel{\scriptsize{$(cut)$} }
 \BinaryInfC{$\tGamma, \tSigma \Rightarrow \tDelta, \tLambda$}
 \DisplayProof
 \vspace{5pt}
\caption{The system $\mathbf{CFL_e}$\label{tableFLe}}
\end{table} 
The calculus $\mathbf{MALL}$ is defined over the language $\mathcal{L}_{\varnothing}^b$ as the calculus consisting of all the axioms and rules in Table \ref{tableFLe} plus the following axioms:
\begin{center}
$\tGamma \Rightarrow \top, \tDelta$
 \quad \quad \quad \quad 
$\tGamma, \bot \Rightarrow \tDelta$
\end{center}
Observe that the rules
\begin{center}
 \AxiomC{$\tGamma, p \Rightarrow \tDelta$}
 \AxiomC{$\tSigma, q \Rightarrow \tLambda$}
 \RightLabel{\scriptsize{$(L+)$}} 
 \BinaryInfC{$\tGamma, \tSigma , p + q \Rightarrow \tDelta, \tLambda$}
 \DisplayProof
 \quad \quad \quad \quad 
 \AxiomC{$\tGamma\Rightarrow p, q, \tDelta$}
 \RightLabel{\scriptsize{$(R+)$}} 
 \UnaryInfC{$\tGamma \Rightarrow p + q, \tDelta$}
 \DisplayProof
\end{center}
are provable in both $\mathbf{CFL_e}$ and $\mathbf{MALL}$, i.e., for any instance of the rule, the conclusion is provable from the set of the premises in $\mathbf{CFL_e}$ and $\mathbf{MALL}$. The calculus
$\mathbf{FL_e}$ over $\mathcal{L}_{\varnothing}^u$ (resp. $\mathbf{IMALL}$ over $\mathcal{L}_{\varnothing}^b$) is  the single-conclusion version of $\mathbf{CFL_e}$ (resp. $\mathbf{MALL}$). 
Now, consider the structural rules:
\begin{center}
Weakening rules: \ \  
 \AxiomC{$\tGamma \Rightarrow \tDelta$}
 \RightLabel{\scriptsize{$(L w)$} }
\UnaryInfC{$\tGamma, p \Rightarrow \tDelta$}
 \DisplayProof
 \quad \quad 
 \AxiomC{$\tGamma \Rightarrow \tDelta$}
 \RightLabel{\scriptsize{$(R w)$} }
 \UnaryInfC{$\tGamma \Rightarrow p, \tDelta$}
 \DisplayProof
 \vspace{7pt}\\
Contraction rules:  \ \ 
 \AxiomC{$\tGamma, p, p \Rightarrow \tDelta$}
 \RightLabel{\scriptsize{$(L c)$} }
\UnaryInfC{$\tGamma, p \Rightarrow \tDelta$}
 \DisplayProof
\quad \quad 
 \AxiomC{$\tGamma\Rightarrow  p, p, \tDelta$}
 \RightLabel{\scriptsize{$(R c)$} }
 \UnaryInfC{$\tGamma \Rightarrow p, \tDelta$}
 \DisplayProof
\end{center}
Adding some or all of these rules to the calculus $\mathbf{CFL_e}$ or their single-conclusion versions to $\mathbf{FL_e}$, we get new calculi over $\mathcal{L}_{\varnothing}^u$:
 $\mathbf{CFL_{ew}}=\mathbf{CFL_{e}}+\{Lw, Rw\}$, $\mathbf{CFL_{ec}}=\mathbf{CFL_{e}}+\{Lc, Rc\}$,  $\mathbf{CFL_{ewc}}=\mathbf{CFL_{ew}}+\{Lc, Rc\}$, 
     $\mathbf{FL_{ew}}=\mathbf{FL_{e}}+\{Lw, Rw\}$, $\mathbf{FL_{ec}}=\mathbf{FL_{e}}+\{Lc\}$, and  $\mathbf{FL_{ewc}}=\mathbf{FL_{ew}}+\{Lc\}$.
Note that
$(R c)$ is only allowed in the multi-conclusion calculi as it has no single-conclusion version. In a similar fashion, we can define the calculi $\mathbf{AMALL}=\mathbf{MALL}+\{Lw, Rw\}$ and  $\mathbf{RMALL}=\mathbf{MALL}+\{Lc, Rc\}$, $\mathbf{AIMALL}=\mathbf{IMALL}+\{Lw, Rw\}$ and  $\mathbf{RIMALL}=\mathbf{IMALL}+\{Lc\}$,
over the language $\mathcal{L}^b_{\varnothing}$.
Observe that in the presence of the weakening rules, $\top$  (resp. $ \bot)$ is provably equivalent to $1$ (resp. $ 0)$, and in the presence of both weakening and contraction rules, the formulas $\phi \wedge \psi$ and $\phi * \psi$ are provably equivalent. 
Hence, the calculus $\mathbf{CFL_{ewc}}$ (resp. $\mathbf{FL_{ewc}}$) is essentially equivalent to the usual calculus $\mathbf{LK}$ (resp. $\mathbf{LJ}$) for the classical (resp. intuitionistic) logic $\CPC$ (resp. $ \IPC)$. As a consequence, we pretend that $\mathbf{LK}$, $\mathbf{LJ}$, $\CPC$, and $\IPC$ are defined over the language of substructural logics.

The second interesting setting is when $I=\{\alpha\}$, i.e., the languages $\mathcal{L}^u_{\{\alpha\}}$ and $\mathcal{L}^b_{\{\alpha\}}$ have only one modality that we  denote by `$!$'. Consider the rules:  
\begin{center}  
\AxiomC{$! \tGamma \Rightarrow p$} \RightLabel{\scriptsize{$(R!)$}} 
\UnaryInfC{$! \tGamma \Rightarrow ! p$} 
\DisplayProof \ \  
\AxiomC{$\tGamma, p \Rightarrow \tDelta$} 
\RightLabel{\scriptsize{$(L!)$}} \UnaryInfC{$\tGamma, ! p \Rightarrow \tDelta$} 
\DisplayProof \ \  
\AxiomC{$ \tGamma \Rightarrow \tDelta$} 
\RightLabel{\scriptsize{$(W!)$}} \UnaryInfC{$ \tGamma, ! p  \Rightarrow \tDelta$} 
\DisplayProof\ \  
\AxiomC{$ \tGamma, ! p , ! p \Rightarrow \tDelta$} 
\RightLabel{\scriptsize{$(C!)$}} \UnaryInfC{$\tGamma, ! p \Rightarrow \tDelta$} 
\DisplayProof
\end{center}
Define the linear calculus $\mathbf{CLL}$ (resp. $\mathbf{ILL}$) as $\mathbf{MALL}$ (resp. $\mathbf{IMALL}$) over $\mathcal{L}^b_{\{\alpha\}}$ plus (resp. the single-conclusion version of) the above rules. Define the \emph{affine} and \emph{relevant} versions of the linear calculi, by adding the weakening and contraction rules, respectively. For instance, define $\mathbf{ALL}=\mathbf{CLL}+\{Lw, Rw\}$ and  $\mathbf{RLL}=\mathbf{CLL}+\{Lc, Rc\}$. The calculus for the \textit{Elementary Linear Logic} \cite{gu,elaine} is $\mathbf{ELL}=\mathbf{MALL}+ \{W!, C! \}$ plus the rules \AxiomC{$\tGamma \Rightarrow p$} \UnaryInfC{$! \tGamma \Rightarrow ! p$} \DisplayProof and \AxiomC{$\tGamma \Rightarrow$} \UnaryInfC{$! \tGamma \Rightarrow$} \DisplayProof.

Let $G$ be a calculus that has the cut rule. It enjoys the \emph{cut elimination}, if any sequent provable in $G$ has a proof in $G$ without using the instances of the cut rule. 
It is proved that the above-mentioned calculi enjoy the cut elimination \cite{Dosenjadid,ono,Gentzen,LLL,affine,kanovich,kiriyama,Ono90,onoketab,OnoKomori,tro92}:

\begin{thm}\label{thm: cut mardom}
The following sequent calculi enjoy the cut elimination:
 \begin{itemize}
     \item 
 $\mathbf{FL_e}$, $\mathbf{FL_{ew}}$, $\mathbf{FL_{ec}}$, 
 $\mathbf{CFL_e}$,   $\mathbf{CFL_{ew}}$,
 $\mathbf{CFL_{ec}}$,
 $\mathbf{LJ}$, 
 $\mathbf{LK}$,
  \item  
 $\mathbf{IMALL}$, $\mathbf{AIMALL}$, $\mathbf{RIMALL}$, 
 $\mathbf{MALL}$,   $\mathbf{AMALL}$, $\mathbf{RMALL}$,
\item 
\cite[Prposition 4]{dosen}  $\mathbf{X}+\{R!, L!\}$, $\mathbf{CFL_e}+\{R!, L!, C!\}$, $\mathbf{CFL_{ew}}+\{R!, L!, C!\}$,
where $\mathbf{X} \in \{\mathbf{CFL_e}, \mathbf{CFL_{ew}},  \mathbf{CFL_{ec}}, \mathbf{LK}\}$,
 \item 
$\mathbf{ILL}$,  $\mathbf{CLL}$,
$\mathbf{ALL}$, $\mathbf{RLL}$, 
$\mathbf{ELL}$.
  \end{itemize}
 \end{thm}

\section{Substructural logics with multimodalities} 
In linear logic, the modality is a machinery to encode non-linearity, allowing an unlimited, yet controlled, access to the resources. Such a machinery can also be used to assign specific locations to the multisets of formulas in proofs to have more expressive proof systems and hence more expressive modes of programming. Keeping this in mind, and as the choice of exponentials is not canonical in linear logic \cite{danos}, one can extend  linear logic with several modal operators, called the \emph{subexponentials}  \cite{nigam,olarte}. These modalities are weaker than the usual $\mathsf{S4}$-type modality in linear logic as they may lack contraction and/or weakening. We can weaken the modalities  even further to capture different modal types (e.g. a $\mathsf{K}$-type or a $\mathsf{D}$-type) of reasoning over a substructural or a fuzzy base. 

In this section, we recall the multimodal substructural logics and their calculi introduced in \cite{elaine} as a general setting for almost all combinations of different types of modalities with basic substructural logics. These calculi use an arbitrary large set $I$ of modalities and hence vastly generalize the calculi presented in Subsection \ref{subsec: 2.1}. Even when $I$ is a singleton, this section provides cut-free calculi for the combinations of substructural logics and weaker forms of modalities than the $\mathsf{S4}$-type modalities in linear logic.
\begin{dfn}\label{Dfn: SDML}
A triple $(I, \preccurlyeq, F)$ is a \emph{simply dependent multimodal logical system} (SDML), where $I$ is a set of indices, $\preccurlyeq$ is a transitive reflexive order on $I$, 
and $F:I \to 2^{\{D, T, 4, C, W\}}$, where $D, T, 4, C, W$  are symbols whose intended meanings become clear in Definition \ref{Dfn: calculus SDML}. An SDML is \emph{suitable} when for each $\alpha, \beta \in I$ that $\alpha \preccurlyeq \beta$, if $4 \in F(\alpha)$ then $4 \in F(\beta)$ and if $C \in F(\alpha)$ then $C \in F(\beta)$. Denote  $\{\beta \in I \mid \alpha \preccurlyeq \beta\}$ by $\uparrow$($\alpha$) and  $\{\beta \in I \mid \alpha \preccurlyeq \beta \; \text{and} \; 4  \in F(\beta)\}$ by $\uparrow^4$($\alpha$).
\end{dfn}
 
\begin{table}[t]
\begin{center}
\AxiomC{$\tGamma_\alpha, \tGamma_{\beta_1}, \dots, \tGamma_{\beta_k}, !^{\gamma_1} \tSigma_{\gamma_1}, \dots, !^{\gamma_m} \tSigma_{\gamma_m} \Rightarrow p$} \RightLabel{\scriptsize{$\ruleK$}} \UnaryInfC{$!^\alpha \tGamma_\alpha, !^{\beta_1} \tGamma_{\beta_1}, \dots, !^{\beta_k} \tGamma_{\beta_k}, !^{\gamma_1} \tSigma_{\gamma_1}, \dots, !^{\gamma_m} \tSigma_{\gamma_m} \Rightarrow !^\alpha p$} \DisplayProof \\  \vspace{5pt}
\AxiomC{$\tGamma_\alpha, \tGamma_{\beta_1}, \dots, \tGamma_{\beta_k}, !^{\gamma_1} \tSigma_{\gamma_1}, \dots, !^{\gamma_m} \tSigma_{\gamma_m} \Rightarrow \tDelta$} \RightLabel{\scriptsize{$\ruleD$}} \UnaryInfC{$!^\alpha \tGamma_\alpha, !^{\beta_1} \tGamma_{\beta_1}, \dots, !^{\beta_k} \tGamma_{\beta_k}, !^{\gamma_1} \tSigma_{\gamma_1}, \dots, !^{\gamma_m} \tSigma_{\gamma_m} \Rightarrow !^\alpha \tDelta$} \DisplayProof \\  \vspace{5pt}
\AxiomC{$\tGamma , p \Rightarrow \tDelta$}
\RightLabel{\scriptsize{$\mathsf{T}_\alpha$}}
 \UnaryInfC{$\tGamma , !^\alpha p \Rightarrow \tDelta$}
 \DisplayProof
\quad
\AxiomC{$\tGamma \Rightarrow \tDelta$}
\RightLabel{\scriptsize{$\mathsf{W}!^\alpha$}}
\UnaryInfC{$\tGamma, !^\alpha p \Rightarrow \tDelta$}
 \DisplayProof \quad
\AxiomC{$ \tGamma, !^\alpha p , !^\alpha p \Rightarrow \tDelta$}
\RightLabel{\scriptsize{$\mathsf{C}!^\alpha$}}
 \UnaryInfC{$\tGamma, !^\alpha p \Rightarrow \tDelta$}
 \DisplayProof\\ 
\vspace{5pt}
\caption{\small{$J=\{\beta_1, \dots, \beta_k\} \subseteq I \setminus \{\alpha\}$ and $L=\{\gamma_1, \dots, \gamma_m\} \subseteq I$ in $\ruleK$ and $\ruleD$ are (possibly empty) finite sets, and the multiset variables $\tGamma_\alpha, \tGamma_{\beta_1}, \dots, \tGamma_{\beta_k},  \tSigma_{\gamma_1}, \dots, \tSigma_{\gamma_m}$ are pairwise distinct. We assume $|\Delta| \leq 1$ in the instances of $\ruleD$ substituting $\tDelta$ by the multiset $\Delta$. }}\label{tableModalAxiom}
\end{center}
\end{table}
Following \cite{elaine}, we  assign a calculus to a suitable SDML:
\begin{dfn}\label{Dfn: calculus SDML}
For a suitable SDML $(I, \preccurlyeq, F)$, the multi-conclusion sequent calculus $\mathbf{CG}_{(I, \preccurlyeq, F)}$ over $\mathcal{L}_I^b$ (resp. $\mathcal{L}_I^{u}$) is the calculus consisting of
the axioms and rules of $\mathbf{MALL}$ (resp. $\mathbf{CFL_e}$) together with the following rules from  Table \ref{tableModalAxiom}, for any  $\alpha \in I$:
\small\begin{equation*}
  \begin{cases}
     \text{$\ruleK$ for any finite $J$ and $L$ that $J \subseteq \uparrow$($\alpha$)$\setminus \{\alpha\}$ and $L \subseteq \uparrow^4$($\alpha$)} &  \text{always} \\
     \text{$\ruleD$ for any finite $J$ and $L$ that $J \subseteq \uparrow$($\alpha$)$\setminus \{\alpha\}$ and $L \subseteq \uparrow^4$($\alpha$)} & \text{if $D \in F(\alpha)$} \\
    \text{$\mathsf{T}_\beta$ for any $\beta \in \uparrow$($\alpha$)} & \text{if $T \in F(\alpha)$} \\
    \text{$\mathsf{C}!^{\alpha}$} & \text{if $C \in F(\alpha)$} \\
     \text{$\mathsf{W}!^{\alpha}$} & \text{if $W \in F(\alpha)$}
  \end{cases}
\end{equation*}
\normalsize The single-conclusion version of $\mathbf{CG}_{(I, \preccurlyeq, F)}$ over 
$\mathcal{L}_I^b$ (resp. $\mathcal{L}_I^{u}$), denoted by $\mathbf{G}_{(I, \preccurlyeq, F)}$, is defined similarly, replacing $\mathbf{MALL}$ by $\mathbf{IMALL}$ (resp. $\mathbf{CFL_e}$ by $\mathbf{FL_e}$) and using the single-conclusion version of its other rules.
\end{dfn}

\begin{rem}\label{rem: D}
The rule $\ruleD$ is a combination of the rule $\ruleK$ and:
\begin{center}
\begin{tabular}{c}
\AxiomC{$\tGamma_\alpha, \tGamma_{\beta_1}, \dots, \tGamma_{\beta_k}, !^{\gamma_1} \tSigma_{\gamma_1}, \dots, !^{\gamma_m} \tSigma_{\gamma_m} \Rightarrow $} 
\UnaryInfC{$!^\alpha \tGamma_\alpha, !^{\beta_1} \tGamma_{\beta_1}, \dots, !^{\beta_k} \tGamma_{\beta_k}, !^{\gamma_1} \tSigma_{\gamma_1}, \dots, !^{\gamma_m} \tSigma_{\gamma_m} \Rightarrow $} 
\DisplayProof
\end{tabular}
\end{center}
where $J=\{\beta_1, \dots, \beta_k\}$ and $L=\{\gamma_1, \dots, \gamma_m\}$. In the modal and linear logic literature and even in \cite{elaine}, the above rule is called $(D)$. However, as it never appears without its $\ruleK$ counterpart, we prefer to merge them in one rule. 
\end{rem}

\begin{exam}\label{exm: logics}
Trivial examples of the logics of $\mathbf{G}_{(I, \preccurlyeq, F)}$ and $\mathbf{CG}_{(I, \preccurlyeq, F)}$ over $\mathcal{L}_I^u$ (resp. $\mathcal{L}_I^b$) are $\mathsf{FL_e}$ and $\mathsf{CFL_e}$ (resp. $\mathsf{IMALL}$ and $\mathsf{MALL}$), when $I=\emptyset$. For the first non-trivial example, let $I=\{\alpha\}$ and note that $F(\alpha) \subseteq \{D, T, 4, C, W\}$. Let us explain the form of the rule $\ruleK$ when $I=\{\alpha\}$. There are two cases: if $4 \in F(\alpha)$ then $J=\emptyset$ and $L \subseteq I$ and if $4 \notin F(\alpha)$ then $J=L=\emptyset$.
Therefore, if $4 \notin F(\alpha)$ (resp. $4 \in F(\alpha)$ and $L=I$), the rule $\ruleK$ is the following left (resp. right) rule:
\begin{center}
    \begin{tabular}{c c}
\AxiomC{$\tGamma_\alpha \Rightarrow p$}
\UnaryInfC{$!^\alpha \tGamma_\alpha \Rightarrow !^\alpha p$}
\DisplayProof
& \hspace{15pt}
\AxiomC{$\tGamma_\alpha, !^{\alpha} \tSigma_{\alpha} \Rightarrow p$}
\UnaryInfC{$!^\alpha \tGamma_\alpha, !^{\alpha} \tSigma_\alpha \Rightarrow !^\alpha p$}
\DisplayProof
    \end{tabular}
\end{center} 
Denoting the unique $!^\alpha$ in the language by $\Box$, the above left (resp. right) rule is the rule $(K)$ (resp. resembles the rule $(K4)$) in the modal logic literature:
\footnotesize 
\begin{center}
\scalebox{1.1}{\begin{tabular}{ c c c c}
\AxiomC{$\tGamma \Rightarrow p$}
\RightLabel{\scriptsize{($K$)}}
 \UnaryInfC{$\Box \tGamma \Rightarrow \Box p$}
 \DisplayProof \hspace{3pt}  
 &
 \AxiomC{$\tGamma , \Box \tSigma \Rightarrow p$} 
\RightLabel{\scriptsize{($K4$)}}
 \UnaryInfC{$\Box \tGamma , \Box \tSigma \Rightarrow \Box p$}
 \DisplayProof \quad \quad \quad 
\end{tabular}}
\end{center}
\normalsize
Hence, one can conclude that $\ruleK$ is a generalization of $(K)$ and $(K4)$. The other rules in Table \ref{tableModalAxiom} are the following familiar rules (see Remark \ref{rem: D}):\vspace{-5pt}
\footnotesize \begin{center}
\scalebox{1.1}{\begin{tabular}{ c c c c}
  \AxiomC{$\tGamma \Rightarrow \tDelta $}
  \RightLabel{\scriptsize{($D$)}}
 \UnaryInfC{$\Box \tGamma \Rightarrow \Box \tDelta$}
 \DisplayProof \hspace{3pt}  &  \AxiomC{$\tGamma, \Box \tSigma \Rightarrow \tDelta$} 
  \RightLabel{\scriptsize{($D4$)}}
 \UnaryInfC{$\Box \tGamma , \Box \tSigma \Rightarrow \Box \tDelta$}
 \DisplayProof
 \end{tabular}}
 \end{center}
 \vspace{-20pt}
 \begin{center}
 \scalebox{1.1}{\begin{tabular}{c c c}
\AxiomC{$\tGamma , p \Rightarrow \tDelta$}
\RightLabel{\scriptsize{($T$)}}
 \UnaryInfC{$\tGamma , \Box p \Rightarrow \tDelta$}
 \DisplayProof \hspace{5pt} & \AxiomC{$\tGamma \Rightarrow \tDelta$}
  \RightLabel{\scriptsize{($W\Box$)}}
\UnaryInfC{$\tGamma, \Box p \Rightarrow \tDelta$}
 \DisplayProof \hspace{5pt} & \AxiomC{$ \tGamma, \Box p, \Box p \Rightarrow \tDelta$}
  \RightLabel{\scriptsize{($C\Box$)}}
 \UnaryInfC{$\tGamma, \Box p \Rightarrow \tDelta$}
 \DisplayProof  
\end{tabular}}
\label{tableModalAxiom2}
\end{center} \vspace{-5pt}
\normalsize
Thus,  for any $X \in \{\mathsf{MALL}, \mathsf{IMALL}, \mathsf{FL_e}, \mathsf{CFL_e}\}$ and $Y \subseteq \{W\Box, C\Box, T\}$, the modal substructural logics $X+(K)+ Y$, $X+(D)+ Y$, $X+(K4)+ Y$ and $X+(D4)+ Y$ 
are examples of the logics of the calculi in Definition \ref{Dfn: calculus SDML}. These logics cover many combinations of the weaker forms of modalities, like $K$-type or $D$-type modalities, with substructural logics. These include: 
\begin{enumerate}
\item 
$\mathsf{CLL}$ is the logic of  $\mathbf{CG}_{(I, \preccurlyeq, F)}$ over $\mathcal{L}_I^b$, if  $F(\alpha)=\{4, T, C, W\}$,
\item 
$\mathsf{ILL}$ is the logic of $\mathbf{G}_{(I, \preccurlyeq, F)}$ over $\mathcal{L}_I^b$, if $F(\alpha)=\{4, T, C, W\}$,
\item 
$\mathsf{ELL}$ is the logic of 
$\mathbf{CG}_{(I, \preccurlyeq, F)}$ over $\mathcal{L}_I^b$, if  
$F(\alpha)=\{D, C, W\}$;
\item 
logic of $\mathbf{CFL_e}+\{R!, L!\}$ (resp. $\mathbf{CFL_e}+\{R!, L!, C!\}$) in  \cite{dosen} is 
the logic of $\mathbf{G}_{(I, \preccurlyeq, F)}$ over $\mathcal{L}_I^u$, if
$F(\alpha)=\{T, 4\}$ (resp. $F(\alpha)=\{T, 4, C\}$). 
\end{enumerate}
In the above observations, we used the fact that in the presence of $(L!)$, or equivalently $(\mathsf{T}_\alpha)$, the rules $(R!)$ and $\mathsf{K(4)}_\alpha^{\varnothing, \{\alpha\}}$ are provably equivalent.
\end{exam}

\begin{exam}
Let $I=\{\alpha, \beta\}$, $\alpha \preccurlyeq \beta$, $F(\alpha)=\{C, W\}$ and $F(\beta)=\{C, 4\}$. Then, $(I, \preccurlyeq, F)$ is a suitable SDML, $\uparrow$($\alpha$) $= \{\alpha, \beta\}$, $\uparrow^4$($\alpha$) $=\{\beta\}$, $\uparrow$($\beta$) =$\uparrow^4$($\beta$) $=\{\beta\}$, and $\mathbf{CG}_{(I, \preccurlyeq, F)}$ over $\mathcal{L}_I^b$ is the extension of $\mathbf{MALL}$ with the rules 
$
C!^\alpha, W!^\alpha, C!^\beta, \mathsf{K(4)}^{J_1,L_1}_\alpha, \mathsf{K(4)}^{J_2,L_2}_\beta
$
for any finite $J_1 \subseteq \uparrow$($\alpha$)$\setminus \{\alpha\}$, $L_1 \subseteq \uparrow^4$($\alpha$), $J_2 \subseteq \uparrow$($\beta$)$\setminus \{\beta\}$, and $L_2 \subseteq \uparrow^4$($\beta$). For instance, for $J_1=L_1=L_2=\{\beta\}$ and $J_2=\emptyset$ the following rules are present in the calculus:
\begin{center}
    \begin{tabular}{c c}
\AxiomC{$\tGamma_\alpha, \tGamma_\beta, !^{\beta} \tSigma_\beta \Rightarrow p$}
\RightLabel{\scriptsize{$\mathsf{K(4)}^{J_1, L_1}_\alpha$}}
\UnaryInfC{$!^\alpha \tGamma_\alpha, !^\beta \tGamma_\beta, !^{\beta} \tSigma_\beta \Rightarrow !^\alpha p$}
\DisplayProof
& \hspace{10pt}
\AxiomC{$\tGamma_\beta, !^{\beta} \tSigma_\beta \Rightarrow p$}
\RightLabel{\scriptsize{$\mathsf{K(4)}^{J_2, L_2}_\beta$}}
\UnaryInfC{$!^\beta \tGamma_\beta, !^{\beta} \tSigma_\beta \Rightarrow !^\beta p$}
\DisplayProof
    \end{tabular}
\end{center}
\end{exam}
\subsection{Cut elimination}
In this subsection, we prove the sequent calculi $\mathbf{CG}_{(I, \preccurlyeq, F)}$ and $\mathbf{G}_{(I, \preccurlyeq, F)}$, both over $\mathcal{L}_I^b$ and $\mathcal{L}_I^{u}$ enjoy cut elimination. To that end, we first present a slightly modified form of the original version of $\mathbf{CG}_{(I, \preccurlyeq, F)}$ over $\mathcal{L}^b_I$ introduced in \cite{elaine}.  Then, we use the cut elimination theorem for this calculus \cite[Theorem 3.1]{elaine}, to establish the cut elimination for $\mathbf{CG}_{(I, \preccurlyeq, F)}$ over $\mathcal{L}^b_I$. Finally, we reduce the cut elimination theorem for the other three calculi to the one we proved. As the calculus in \cite{elaine} is one-sided and defined over a different language from ours, we must temporarily extend our notions of a language and sequent calculus. This modification is only limited to this subsection.

To pursue this strategy, 
define the language  $\mathcal{L}_{\mathbf{H}}$ as the extension of $\mathcal{L}_I^b \setminus \{\to\}$ by the duals of the atoms, denoted $p^{\bot}$, along with the connectives $+$ and $?^{\alpha}$, for any $\alpha \in I$, i.e., $\mathcal{L}_{\mathbf{H}}=\{0,1, \bot, \top, \wedge, \vee, * ,+\} \cup \{!^{\alpha}, ?^{\alpha} \mid \alpha \in I\}$. Define the dualization operator $\theta \mapsto \theta^\bot$ on $\mathcal{L}_{\mathbf{H}}$ as usual, e.g.,
$(\phi * \psi)^{\bot} :=\phi^\bot + \psi^{\bot}$,  $(\phi \vee \psi)^{\bot}:=\phi^{\bot} \wedge \psi^{\bot}$, and $(!^{\alpha} \phi)^{\bot} := ?^{\alpha}\phi^\bot$. Now, define the calculus $\mathbf{H}_{(I, \preccurlyeq, F)}$ as the usual one-sided version of $\mathbf{MALL}$ augmented with the rules in Table \ref{tableModalAxiomOriginal} in the same way as in Definition \ref{Dfn: calculus SDML}. 
\begin{table}[t]
\begin{center}
\AxiomC{$\tGamma_\alpha, \tGamma_{\beta_1}, \dots, \tGamma_{\beta_k}, ?^{\gamma_1} \tSigma_{\gamma_1}, \dots, ?^{\gamma_m} \tSigma_{\gamma_m}, p$}
\RightLabel{\scriptsize{1-$\mathsf{K(4)}^{J, L}_\alpha$}}\UnaryInfC{$?^\alpha \tGamma_\alpha, ?^{\beta_1} \tGamma_{\beta_1}, \dots, ?^{\beta_k} \tGamma_{\beta_k}, ?^{\gamma_1} \tSigma_{\gamma_1}, \dots, ?^{\gamma_m} \tSigma_{\gamma_m}, !^\alpha p$} \DisplayProof \\  \vspace{3pt}
\AxiomC{$\tGamma_\alpha, \tGamma_{\beta_1}, \dots, \tGamma_{\beta_k}, ?^{\gamma_1} \tSigma_{\gamma_1}, \dots, ?^{\gamma_m} \tSigma_{\gamma_m}, \tDelta$} 
\RightLabel{\scriptsize{1-$\mathsf{D(4)}^{J, L}_\alpha$}}
\UnaryInfC{$?^\alpha \tGamma_\alpha, ?^{\beta_1} \tGamma_{\beta_1}, \dots, ?^{\beta_k} \tGamma_{\beta_k}, ?^{\gamma_1} \tSigma_{\gamma_1}, \dots, ?^{\gamma_m} \tSigma_{\gamma_m}, !^\alpha \tDelta$} \DisplayProof \\  \vspace{3pt}
\AxiomC{$\tGamma , p$}
\RightLabel{\scriptsize{$\mathsf{T}?^\alpha$}}
 \UnaryInfC{$\tGamma , ?^\alpha p$}
 \DisplayProof
\quad
\AxiomC{$\tGamma$}
\RightLabel{\scriptsize{$\mathsf{W}?^\alpha$}}
\UnaryInfC{$\tGamma, ?^\alpha p$}
 \DisplayProof \quad
\AxiomC{$ \tGamma, ?^\alpha p , ?^\alpha p$}
\RightLabel{\scriptsize{$\mathsf{C}?^\alpha$}}
 \UnaryInfC{$\tGamma, ?^\alpha p $}
 \DisplayProof
\caption{\small{$J=\{\beta_1, \dots, \beta_k\} \subseteq I \setminus \{\alpha\}$ and $L=\{\gamma_1, \dots, \gamma_m\} \subseteq I$ in 1-$\ruleK$ and 1-$\ruleD$ are (possibly empty) finite sets, and the multiset variables $\tGamma_\alpha, \tGamma_{\beta_1}, \dots, \tGamma_{\beta_k},  \tSigma_{\gamma_1}, \dots, \tSigma_{\gamma_m}$ are pairwise distinct. We assume $|\Delta| \leq 1$ in the instances of 1-$\ruleD$ substituting $\tDelta$ by the multiset $\Delta$. The ``1" in the first two rules indicates the one-sided form of the rules.}}\label{tableModalAxiomOriginal}
\end{center}
\end{table}
The system $\mathbf{H}_{(I, \preccurlyeq, F)}$ is essentially equivalent to the one introduced in \cite{elaine}. However, we slightly changed the presentation of its rules to align them with ours. This modification is immaterial for our purpose of cut elimination as any cut-free proof in $\mathbf{H}_{(I, \preccurlyeq, F)}$ can be simulated by a cut-free proof in the system in \cite{elaine} and vice versa.
To see why, note that the  original version of the rule 1-$\ruleK$, called $\mathsf{K4}_{\alpha}$ in \cite{elaine}, has the form:
\begin{center}
\AxiomC{$ \tGamma_{\beta_1}, \dots, \tGamma_{\beta_k}, ?^{\gamma_1} \tSigma_{\gamma_1}, \dots, ?^{\gamma_m} \tSigma_{\gamma_m}, p$}
\RightLabel{\scriptsize{$\mathsf{K4}_\alpha$}}\UnaryInfC{$?^{\beta_1} \tGamma_{\beta_1}, \dots, ?^{\beta_k} \tGamma_{\beta_k}, ?^{\gamma_1} \tSigma_{\gamma_1}, \dots, ?^{\gamma_m} \tSigma_{\gamma_m}, !^\alpha p$} 
\DisplayProof 
\end{center}
for any $\{\beta_1, \ldots, \beta_k\} \subseteq \uparrow\!\!(\alpha)$ and  $\{\gamma_1, \ldots, \gamma_k\} \subseteq \uparrow^4\!\!(\alpha)$. The difference between $\mathsf{K4}_\alpha$ and 1-$\ruleK$ is that the former has no specified place for $\tGamma_{\alpha}$. Clearly, the family of the rules $\mathsf{K4}_{\alpha}$ includes our 1-$\ruleK$ rules for any $J \subseteq \uparrow\!\!(\alpha) \setminus \{\alpha\}$ and $L \subseteq \uparrow^4\!\!(\alpha)$. For the converse, as one can always substitute the empty multiset for $\tGamma_{\alpha}$ in 1-$\ruleK$, our version also simulates $\mathsf{K4}_{\alpha}$. Similarly, the family of our 1-$\ruleD$ rules are equivalent to the family of the rules:
\begin{center}
\AxiomC{$ \tGamma_{\beta_1}, \dots, \tGamma_{\beta_k}, ?^{\gamma_1} \tSigma_{\gamma_1}, \dots, ?^{\gamma_m} \tSigma_{\gamma_m}, \tDelta$}
\UnaryInfC{$?^{\beta_1} \tGamma_{\beta_1}, \dots, ?^{\beta_k} \tGamma_{\beta_k}, ?^{\gamma_1} \tSigma_{\gamma_1}, \dots, ?^{\gamma_m} \tSigma_{\gamma_m}, !^\alpha \tDelta$} 
\DisplayProof 
\end{center}
where $\{\beta_1, \ldots, \beta_k\} \subseteq \uparrow\!\!(\alpha)$ and  $\{\gamma_1, \ldots, \gamma_k\} \subseteq \uparrow^4\!\!(\alpha)$.
These rules are clearly a combination of the rules $\mathsf{D}_\alpha$ in \cite{elaine}
\begin{center}
\AxiomC{$ \tGamma_{\beta_1}, \dots, \tGamma_{\beta_k}, ?^{\gamma_1} \tSigma_{\gamma_1}, \dots, ?^{\gamma_m} \tSigma_{\gamma_m}$}
\RightLabel{\scriptsize{$\mathsf{D}_\alpha$}}
\UnaryInfC{$?^{\beta_1} \tGamma_{\beta_1}, \dots, ?^{\beta_k} \tGamma_{\beta_k}, ?^{\gamma_1} \tSigma_{\gamma_1}, \dots, ?^{\gamma_m} \tSigma_{\gamma_m}$} 
\DisplayProof 
\end{center}
and $\mathsf{K4}_{\alpha}$. Finally, the $T$-type rule in \cite{elaine} is
\begin{center}
\AxiomC{$\tGamma, \tGamma_{\beta_1}, \dots, \tGamma_{\beta_k}$}
\RightLabel{\scriptsize{$\mathsf{T}_\alpha$}}
\UnaryInfC{$\tGamma, ?^{\beta_1} \tGamma_{\beta_1}, \dots, ?^{\beta_k} \tGamma_{\beta_k}$} 
\DisplayProof 
\end{center}
where $\{\beta_1, \ldots, \beta_k\} \subseteq \uparrow\!\!(\alpha)$. Clearly, $\mathsf{T}_\alpha$ is equivalent to a combination of our rules $\mathsf{T}?^{\beta}$, for any $
\beta \in \uparrow\!\!(\alpha)$.

Next, we modify $\mathbf{H}_{(I, \preccurlyeq, F)}$ to have an explicit rule for the implication. 
To this end, extend the language $\mathcal{L}_{\mathbf{H}}$ of $\mathbf{H}_{(I, \preccurlyeq, F)}$ to $\mathcal{L}_{\mathbf{H}}^\to=\mathcal{L}_{\mathbf{H}} \cup \{\to\}$ and the dualization operator $\theta \mapsto \theta^\bot$ to $\mathcal{L}_{\mathbf{H}}^\to$ by defining $(\phi \to \psi)^{\bot}:=\phi * \psi^{\bot}$. Define the calculus $\mathbf{H}^\to_{(I, \preccurlyeq, F)}$ over $\mathcal{L}_{\mathbf{H}}^\to$ as the union of $\mathbf{H}_{(I, \preccurlyeq, F)}$ over $\mathcal{L}^{\to}_{\mathbf{H}}$ and the rule:
\begin{center}
\AxiomC{$\tSigma, p^{\bot}, q$}
\UnaryInfC{$\tSigma, p \to q$} 
\DisplayProof \quad $(\dagger)$
\end{center}
The cut elimination theorem still holds for $\mathbf{H}^\to_{(I, \preccurlyeq, F)}$ over $\mathcal{L}_{\mathbf{H}}^\to$: 

\begin{lem}\label{CutElimForH}
The calculus $\mathbf{H}^\to_{(I, \preccurlyeq, F)}$ over $\mathcal{L}_{\mathbf{H}}^\to$ enjoys the cut elimination.
\end{lem}
\begin{proof}
Define the function $s:\mathcal{L}_{\mathbf{H}}^\to \to \mathcal{L}_{\mathbf{H}}$ recursively as the identity function on $p, p^\bot, 0, 1, \top, \bot$, for $p$ an atom, $(\phi \circ \psi)^s=\phi^s \circ \psi^s$, for $\circ \in \{\wedge, \vee, +, *\}$ and $(\phi \to \psi)^s=(\phi^\bot)^s + \psi^s$. Assume $\mathbf{H}^\to_{(I, \preccurlyeq, F)} \vdash \Gamma$, for a multiset 
$\Gamma$ of formulas in $\mathcal{L}_{\mathbf{H}}^{\to}$. Then, by induction on the structure of the proof, it is easy to see that  $\mathbf{H}_{(I, \preccurlyeq, F)} \vdash
\Gamma^s$. By \cite[Theorem 3.1]{elaine} and the above discussions, there is a cut-free proof of $\Gamma^s$ in $\mathbf{H}_{(I, \preccurlyeq, F)}$.  By an easy induction on the structure of the proof of $\Gamma^s$  in $\mathbf{H}_{(I, \preccurlyeq, F)}$, we provide a cut-free proof for $\Gamma$ in $\mathbf{H}^\to_{(I, \preccurlyeq, F)}$. The non-trivial case is when the last rule used in the proof of $\Gamma^s$ is the rule for $+$:
\begin{center}
    \AxiomC{$\Sigma^s, \phi, \psi$}
    \UnaryInfC{$\Sigma^s, \phi + \psi$}
    \DisplayProof
\end{center}
Then, $\phi+\psi=\theta^s$, where $\theta \in \mathcal{L}^{\to}_{\mathbf{H}}$. Thus, either $\theta=\phi' \to  \psi'$ or $\theta=\phi' + \psi'$, where $\phi',\psi' \in \mathcal{L}^{\to}_{\mathbf{H}}$. We explain the former case, where $\theta^s = (\phi' \to  \psi')^s = (\phi'^\bot)^s +  \psi'^s$. Hence, $\phi=(\phi'^{\bot})^s$ and $\psi=\psi'^s$ and the premise of the rule has the form $\Sigma^s, (\phi'^\bot)^s ,  \psi'^s$. By the induction hypothesis and then applying the rule $(\dagger)$, we get a cut-free proof for 
$\Sigma, \phi' \to \psi'$ in $\mathbf{H}^\to_{(I, \preccurlyeq, F)}$, as required.
\end{proof}

To prove the cut elimination for $\mathbf{CG}_{(I, \preccurlyeq, F)}$ over $\mathcal{L}_I^b$, we employ the usual technique of showing the equivalence between a one-sided calculus and its two-sided counterpart. In this regard, we need the following lemma:

\begin{lem} \label{NegIsInvertible}
Let $(I, \preccurlyeq, F)$ be a suitable SDML. For any multisets $\Sigma$, $\Pi$, and $\Lambda$ over $\mathcal{L}_I^b$:
\begin{itemize}
\item[$(i)$] 
if there is a cut-free proof of $ \Sigma \Rightarrow 0, \Lambda$ in $\mathbf{CG}_{(I, \preccurlyeq, F)}$ over $\mathcal{L}_I^b$, then $ \Sigma \Rightarrow \Lambda$ also has a cut-free proof in the same system.
\item[$(ii)$]
if there is a cut-free proof of $\Sigma \Rightarrow \neg \Pi, \Lambda$ in $\mathbf{CG}_{(I, \preccurlyeq, F)}$ over $\mathcal{L}_I^b$, then $ \Sigma, \Pi \Rightarrow \Lambda$ also has a cut-free proof in the same system.
\end{itemize}  
\end{lem}
\begin{proof}
Use induction on the structure of the proof. The non-trivial case in $(ii)$ is when the last rule is $(R\to)$ and the main formula is in $\neg \Pi=\neg \Pi', \neg \phi$:
\begin{center}
    \AxiomC{$\Sigma, \phi \Rightarrow 0, \neg \Pi', \Lambda$}
\RightLabel{\small$(R \to)$}
\UnaryInfC{$\Sigma \Rightarrow \phi \to 0, \neg \Pi', \Lambda$}
\DisplayProof
\end{center}
By the induction hypothesis, there is a cut-free proof for $\Sigma, \phi, \Pi' \Rightarrow 0, \Lambda$ and by $(i)$ there is a cut-free proof for $\Sigma, \phi, \Pi' \Rightarrow \Lambda$ in the system, as required.
\end{proof}

Define the translation function $t:\mathcal{L}_{\mathbf{H}}^\to \to \mathcal{L}^b_I$ as the identity function on atoms and constants, $(p^\bot)^t=\neg p$, for $p$ an atom, $(\phi \circ \psi)^t=\phi^t \circ \psi^t$, for $\circ \in \{\wedge, \vee, *, \to\}$, $(\phi + \psi)^t=\neg(\neg \phi^t * \neg \psi^t)$, $(!^\alpha \phi)^t= !^\alpha \phi^t$, and $(?^\alpha \phi)^t= \neg !^\alpha \neg \phi^t$.
For a multiset $\Sigma$, we define $\Sigma^t= \{\theta^t \mid \theta \in \Sigma\}$ and $\Sigma^\bot= \{\theta^\bot \mid \theta \in \Sigma\}$.
\begin{lem} \label{thm: elaine}
Let $(I, \preccurlyeq, F)$ be a suitable SDML. Denote $\mathbf{CG}_{(I, \preccurlyeq, F)}$ over $\mathcal{L}_I^b$ by $G$ and $\mathbf{H}^\to_{(I, \preccurlyeq, F)}$ over $\mathcal{L}_{\mathbf{H}}^\to$ by $H$. 
\begin{itemize}
\item[$(i)$]
For any sequent $S=(\Gamma \Rightarrow \Delta)$ over $\mathcal{L}_{I}^b$, if $S$ is provable in $G$, then $\Gamma^{\bot}, \Delta$ is provable in $H$ (with a possible use of cut).
\item[$(ii)$]
For any multisets $\Sigma$ and $\Lambda$ over $\mathcal{L}_{\mathbf{H}}^\to$, if
$\Sigma^{\bot}, \Lambda$ has a cut-free proof in $H$, then $\Sigma^t \Rightarrow \Lambda^t$ has a cut-free proof in $G$. 
\end{itemize}
\end{lem}

\begin{proof}
Part $(i)$ is straightforward. For $(ii)$, we use induction on the structure  
of the cut-free proof in $H$. The only complicated cases are the modal rules in Table \ref{tableModalAxiomOriginal}. We only check the case where the last rule is $(\mathsf{C}?^{\alpha})$:
\begin{center}
\begin{tabular}{c c}
\AxiomC{$\Gamma, ?^{\alpha} \phi, ?^{\alpha} \phi$}
\UnaryInfC{$\Gamma, ?^{\alpha} \phi$}
\DisplayProof
    \end{tabular}
\end{center}
Recall $\Sigma^\bot= \{\theta^\bot \mid \theta \in \Sigma\}$. As $\Gamma, ?^{\alpha} \phi=\Sigma^\bot, \Lambda$, there are two cases, either $?^{\alpha} \phi \in \Sigma^{\bot}$ or $?^{\alpha} \phi \in \Lambda$. In the first case, $\Sigma^\bot=\Gamma_1^\bot, ?^\alpha \phi$ and $\Gamma_2=\Lambda$, where $\Gamma=\Gamma_1^\bot \cup \Gamma_2$. Since
$?^{\alpha} \phi \in \Sigma^{\bot}$, there is a formula $\psi$ such that $?^{\alpha} \phi= (!^\alpha \psi)^\bot$. By the induction hypothesis, there is a cut-free proof of $\Gamma_1^t, !^{\alpha} \psi^t, !^{\alpha} \psi^t \Rightarrow \Gamma_2^t$ in $G$. By $(\mathsf{C}!^{\alpha})$ in $G$, we get $\Gamma_1^t, !^{\alpha} \psi^t\Rightarrow \Gamma_2^t$ in $G$, as required. For the second case, $\Lambda= ?^\alpha \phi, \Gamma_2$, where $\Gamma=\Gamma_1^\bot \cup \Gamma_2$. By the induction hypothesis, there is a cut-free proof in $G$ for $\Gamma_1^t \Rightarrow (?^{\alpha} \phi)^t,  (?^{\alpha} \phi)^t, \Gamma_2^t$, which is $\Gamma_1^t \Rightarrow \neg !^{\alpha} \neg \phi^t,  \neg !^{\alpha} \neg \phi^t,  \Gamma_2^t$. By Lemma \ref{NegIsInvertible}, there is a cut-free proof of $\Gamma_1^t, !^{\alpha} \neg \phi^t, !^{\alpha} \neg \phi^t \Rightarrow \Gamma_2^t$ in $G$. Again, by $(\mathsf{C}!^{\alpha})$ in $G$ and then applying $(0w)$ and $(R\to)$, we get $\Gamma_1^t \Rightarrow  \neg !^{\alpha} \neg \phi^t,  \Gamma_2^t$, as required.
\end{proof}
Recall that any cut-free proof has a nice property known as the \emph{subformula property} meaning that all formulas occurring in the proof of a sequent $\Gamma \Rightarrow \Delta$ are subformulas of the formulas either in $\Gamma$ or $\Delta$.
\begin{cor} \label{cor: elaine}
    If $(I, \preccurlyeq, F)$ is a suitable SDML, then  $\mathbf{CG}_{(I, \preccurlyeq, F)}$ and $\mathbf{G}_{(I, \preccurlyeq, F)}$ over $\mathcal{L}_I^b$ or $\mathcal{L}_I^{u}$ enjoy the cut elimination.
\end{cor}
\begin{proof}
The case for
$\mathbf{CG}_{(I, \preccurlyeq, F)}$ over $\mathcal{L}_I^b$ is a consequence of Lemma \ref{CutElimForH}, Lemma \ref{thm: elaine}, and the easy observation that if $\phi$ is an $\mathcal{L}_I^b$-formula, then $\phi^t=\phi$. 
\item[$\bullet$]
$\mathbf{G}_{(I, \preccurlyeq, F)}$ over $\mathcal{L}_I^b$: If a single-conclusion sequent is provable in $\mathbf{G}_{(I, \preccurlyeq, F)}$, it is also provable in $\mathbf{CG}_{(I, \preccurlyeq, F)}$. Therefore, by the previous part, it has a cut-free proof $\pi$ in $\mathbf{CG}_{(I, \preccurlyeq, F)}$. As the sequent is single-conclusion, the proof $\pi$ is single-conclusion, because going bottom up in $\pi$, the succedents of the sequents cannot increase in size and hence they will have at most one formula. Therefore, the cut-free proof $\pi$ actually lives in $\mathbf{G}_{(I, \preccurlyeq, F)}$.
\item[$\bullet$]
$\mathbf{G}_{(I, \preccurlyeq, F)}$ and $\mathbf{CG}_{(I, \preccurlyeq, F)}$ over $\mathcal{L}_I^{u}$: We only prove the claim for $\mathbf{CG}_{(I, \preccurlyeq, F)}$. If a sequent, not containing $\bot, \top$, is provable in $\mathbf{CG}_{(I, \preccurlyeq, F)}$ over $\mathcal{L}_I^{u}$, then it is also provable in $\mathbf{CG}_{(I, \preccurlyeq, F)}$ over $\mathcal{L}_I^{b}$. By the previous parts, it
has a cut-free proof $\pi$ in $\mathbf{CG}_{(I, \preccurlyeq, F)}$ over $\mathcal{L}_I^{b}$. By the subformula property, $\bot, \top$ are not present in $\pi$. Therefore, the cut-free proof $\pi$ actually lives in $\mathbf{CG}_{(I, \preccurlyeq, F)}$ over $\mathcal{L}_I^{u}$. 
\end{proof}

\section{Semi-analytic Calculi} 
We introduce a class of axioms and rules specified by their special, yet sufficiently general, syntactical form. We call a calculus consisting of these axioms and rules \emph{semi-analytic} and argue that the family of such calculi offers a close approximation of what is considered a ``nice" calculus in the literature. 
\begin{dfn}
A rule is \emph{occurrence-preserving} if it has exactly one main formula and the set of the variables of the active formulas is a subset of the set of the variables of the main formula.
\end{dfn}
Note that this property is weaker than the analyticity property in analytic rules, where formulas in the premises are subformulas of the main formula.

\begin{exam}
The following rule is occurrence-preserving
\small\begin{center}
 \AxiomC{$\{\tPi_j , \bar{\nu}_{js} \Rightarrow \bar{\rho}_{js} \mid 1 \leq j \leq m, 1 \leq s \leq l_j \}$} 
 \AxiomC{$\{\tGamma_i , \bar{\mu}_{ir} \Rightarrow \tDelta_i \mid 1 \leq i \leq n, 1 \leq r \leq k_i \}$}
 \BinaryInfC{$\tPi_1, \dots, \tPi_m, \tGamma_1, \dots, \tGamma_n, \mu \Rightarrow \tDelta_1, \dots, \tDelta_n $}
 \DisplayProof
\end{center}
\normalsize when 
$ \bigcup_{i, r} V(\bar{\mu}_{ir}) \cup \bigcup_{j, s} V(\bar{\nu}_{js}) \cup \bigcup_{j, s} V(\bar{\rho}_{js}) \subseteq V(\mu)$. Concrete examples of occurrence-preserving rules are  the rules in 
Table \ref{tableFLe}, except for the cut rule.
The rule
\begin{center}
 \AxiomC{$\tGamma \Rightarrow p, \tDelta$} 
 \AxiomC{$\tGamma \Rightarrow p \to q, \tDelta$}
 \BinaryInfC{$\tGamma \Rightarrow p \wedge q, \tDelta$}
 \DisplayProof
\end{center}
is an instance of an occurrence-preserving rule that is not analytic, as $p \to q$ is not a subformula of $p \wedge q$.
\end{exam} 

\begin{dfn} \label{semi-analyticRules}
For $1 \leq i \leq n$ and $1 \leq j \leq m$, let $\tGamma_i$'s, $\tDelta_i$'s, and $\tPi_j$'s be pairwise disjoint families of pairwise distinct multiset variables, $\bar{\mu}_{ir}$'s, $\bar{\nu}_{js}$'s, and $\bar{\rho}_{js}$'s be multisets of formulas and $\mu$ a formula, where $1 \leq r \leq k_i$ and $1 \leq s \leq l_j$. It is possible to have the value $0$ for $n$, $m$, $k_i$, and $l_j$ which means that $i$, $j$, $r$, and $s$ range over the empty set, respectively and if there is no fear of confusion, we omit the domain of these indices. A rule is \emph{single-conclusion semi-analytic} if it is occurrence-preserving and has one of the following forms:
\begin{itemize}
\item
\emph{left single-conclusion semi-analytic}:
\end{itemize}
\small\begin{center}
 \AxiomC{$\{\tPi_j , \bar{\nu}_{js} \Rightarrow \bar{\rho}_{js} \mid 1 \leq j \leq m, 1 \leq s \leq l_j \}$} 
 \AxiomC{$\{\tGamma_i , \bar{\mu}_{ir} \Rightarrow \tDelta_i \mid 1 \leq i \leq n, 1 \leq r \leq k_i \}$}
 \BinaryInfC{$\tPi_1, \dots, \tPi_m, \tGamma_1, \dots, \tGamma_n, \mu \Rightarrow \tDelta_1, \dots, \tDelta_n $}
 \DisplayProof
\end{center}
\normalsize where 
$|\bar{\rho}_{js}| \leq 1$, for each $j$ and $s$. Note that in each instance of the rule, at most one of $\tDelta_i$'s can be substituted by a formula and the rest must be empty, as the rule is single-conclusion. Note that if $n=0$, there is no premise of the right branch and the conclusion is of the form $\tPi_1, \dots, \tPi_m, \mu \Rightarrow$. Similarly, if $m=0$, the conclusion has the form $\tGamma_1, \dots, \tGamma_n, \mu \Rightarrow \tDelta_1, \dots, \tDelta_n$.
\begin{itemize}
\item
\emph{right single-conclusion semi-analytic}:
\end{itemize}
\small\begin{center}
 \AxiomC{$\{\tGamma_i , \bar{\mu}_{ir} \Rightarrow \bar{\nu}_{ir} \mid 1 \leq i \leq n, 1 \leq r \leq k_i\}$}
 \UnaryInfC{$\tGamma_1, \dots, \tGamma_n \Rightarrow \mu $}
 \DisplayProof
\end{center}
\normalsize where $|\bar{\nu}_{ir}| \leq 1$, for each $i$ and $r$ and if $n=0$, the conclusion has the form $\Rightarrow \mu$.
A rule is called \emph{multi-conclusion semi-analytic} if it is occurrence-preserving and has one of the following forms:
\begin{itemize}
\item
\emph{left multi-conclusion semi-analytic}:
\end{itemize}
\small \begin{center}
 \AxiomC{$\{ \tGamma_i , \bar{\mu}_{ir} \Rightarrow \bar{\nu}_{ir}, \tDelta_i \mid 1 \leq i \leq n, 1 \leq r \leq k_i \}$}
 \UnaryInfC{$\tGamma_1, \dots, \tGamma_n, \mu \Rightarrow \tDelta_1, \dots, \tDelta_n $}
 \DisplayProof
\end{center}
\normalsize \begin{itemize}
\item
\emph{right multi-conclusion semi-analytic}:
\end{itemize}
\small \begin{center}
 \AxiomC{$\{ \tGamma_i , \bar{\mu}_{ir} \Rightarrow \bar{\nu}_{ir}, \tDelta_i \mid 1 \leq i \leq n, 1 \leq r \leq k_i\}$}
 \UnaryInfC{$\tGamma_1, \dots, \tGamma_n \Rightarrow \mu, \tDelta_1, \dots, \tDelta_n $}
 \DisplayProof
\end{center} 
\normalsize and if $n=0$, the conclusion of the rules has either the form $\mu \Rightarrow$ or $\Rightarrow \mu$. 
\end{dfn}

\begin{rem}
The rules in Definition \ref{semi-analyticRules} are called \emph{semi-analytic}, as their main
condition of occurrence-preservation is a weaker version of the usual analyticity property. Furthermore, recall that the usual sequent calculi for substructural logics have two basic ways to handle the contexts. For the additive connectives, namely the conjunction and  disjunction, the contexts in the premises of the rule must be the same and the conclusion inherits these common contexts. For the multiplicative connectives, i.e., the implication and fusion, the rule combines the contexts of its premises and inserts the combination in its conclusion. Semi-analytic rules allow almost all possible combinations of these two approaches. For instance, in the premises of a right multi-conclusion semi-analytic rule, fixing $i$, the family $\{ \tGamma_i , \bar{\mu}_{ir} \Rightarrow \bar{\nu}_{ir}, \tDelta_i \mid 1 \leq r \leq k_i\}$ shares the common  context $\tGamma_i$ and $\tDelta_i$, while ranging over $i$, the rule combines them to reach $\tGamma_1, \ldots, \tGamma_n$ and $\tDelta_1, \ldots, \tDelta_n$ in the conclusion.
The only exception is the case of left single-conclusion semi-analytic rules.
In these rules, there are two \emph{types} of premises: either the succedent contains at most one formula (the left branch of premises) or one context (the right branch). Between these two branches, we can only combine the contexts in the antecedents and sharing them is not an option. 
Finally,
semi-analytic rules exhaust all possible combinatorial ways of having an occurrence-preserving rule satisfying the following three conditions: 
the multiset variables are free for any multiset substitution, they do not change in the rule application from the premises to the conclusion, and in each single-conclusion (resp. multi-conclusion) rule there are multiset variables in the antecedent (resp. in the antecedent and succedent) of the conclusion of the rule.
This combinatorial exhaustion alongside the occurrence-preservation explains why we believe that semi-analytic rules form a sufficiently general family of rules to approximate the informal notion of a ``nice" rule.
\end{rem}

\begin{exam}\label{ExampleSemi-analyticRules}
A generic example of a left single-conclusion semi-analytic rule is the following:
\begin{center}
 \begin{tabular}{c }
 \AxiomC{$\tPi, \nu^1_{11}, \nu^2_{11} \Rightarrow \rho_{11}$}
 \AxiomC{$\tPi, \nu_{12} \Rightarrow \rho_{12}$}
 \AxiomC{$\tGamma, \mu^1_1, \mu^2_1, \mu^3_1 \Rightarrow \tDelta$}
 \TrinaryInfC{$\tPi, \tGamma, \mu \Rightarrow \tDelta$}
 \DisplayProof
\end{tabular}
\end{center}
where 
\[
V(\mu^1_1) \cup V(\mu^2_1) \cup V(\mu_1^3) \cup V(\nu_{11}^1) \cup V(\nu_{11}^2) \cup V(\nu_{12}) \cup V(\rho_{11}) \cup V(\rho_{12}) \subseteq V(\mu).
\]
The left and middle premises are  in the same family with the context $\tPi$ in the antecedent. Thus, one copy of $\tPi$ appears in the antecedent of the conclusion. A  generic example of a right multi-conclusion semi-analytic rule is:
\begin{center}
 \begin{tabular}{c }
 \AxiomC{$\tGamma_1, \mu_{11} \Rightarrow \nu_{11}, \tDelta_1$}
 \AxiomC{$\tGamma_1, \mu_{12}^1, \mu_{12}^2 \Rightarrow \tDelta_1$}
 \AxiomC{$\tGamma_2, \mu_{2}^1, \mu_{2}^2 \Rightarrow \nu_{2}^1, \nu_{2}^2, \tDelta_2$}
 \TrinaryInfC{$\tGamma_1, \tGamma_2 \Rightarrow \mu, \tDelta_1, \tDelta_2$}
 \DisplayProof
\end{tabular}
\end{center}
where 
\[
V(\mu_{11}) \cup V(\mu^1_{12}) \cup V(\mu^2_{12}) \cup V(\mu^1_{2}) \cup V(\mu^2_{2}) \cup V(\nu_{11}) \cup V(\nu_{2}^1) \cup V(\nu_{2}^2) \subseteq V(\mu).
\] 
Again, the left and middle premises belong to the same family. 
\end{exam}

\begin{exam}
Concrete examples of multi-conclusion semi-analytic rules are the rules in Table \ref{tableFLe} (except the cut rule), the weakening and contraction rules, and the rules $\mathsf{T}_\alpha, \mathsf{W}!^\alpha, \mathsf{C}!^\alpha$ in Table \ref{tableModalAxiom}. The single-conclusion version of these rules are single-conclusion semi-analytic rules.
Concrete non-examples of multi-conclusion semi-analytic rules include the cut rule, as it violates the occurrence-preserving condition, $\ruleK$ and $\ruleD$ in Table \ref{tableModalAxiom}, as the contexts do not remain intact from the premises to the conclusion (some like $\Gamma_j$ and $\Delta$ transform to $!^{\beta_j} \Gamma_j$ and $!^\alpha \Delta$, respectively). Non-examples for single-conclusion semi-analytic rules are provided by the single-conclusion version of the mentioned rules. For other non-examples, consider the following rules in the calculus $\mathbf{KC}$ introduced in \cite{bor} for the logic $\mathsf{KC}=\mathsf{IPC}+\neg p \vee \neg \neg p$: 
\begin{center}
 \begin{tabular}{c c}
 \AxiomC{$\tGamma, p \Rightarrow q, \tDelta$}
 \UnaryInfC{$\tGamma \Rightarrow p \to q, \tDelta$}
 \DisplayProof
 & \ \ 
  \AxiomC{$\tGamma \Rightarrow \tDelta, p$}
   \AxiomC{$q, \tPi \Rightarrow \tLambda$}
 \BinaryInfC{$\tGamma, p \to q, \tPi \Rightarrow  \tDelta, \tLambda$}
 \DisplayProof
\end{tabular}
\end{center}
in which $\tDelta$ must be substituted by multisets of \emph{strictly negative formulas} (not containing all formulas in the language; see \cite{bor}).  
These rules are not semi-analytic as their context $\tDelta$ is not free for arbitrary substitutions of multisets.
\end{exam}

\begin{dfn} \label{Dfn Focused Axioms}
A meta-sequent is a \emph{focused axiom}, if it has one of the forms:
\begin{center}
 $\mu \Rightarrow \mu \quad \quad  \quad \quad \Rightarrow \bar{\nu} \quad \quad  \quad \quad  \bar{\rho} \Rightarrow \quad \quad  \quad \quad \tGamma, \bar{\mu} \Rightarrow \tDelta \quad \quad  \quad \quad  \tGamma \Rightarrow \bar{\mu}, \tDelta$   
\end{center}
where $V(\phi)= V(\psi)$, for any $\phi, \psi \in \bar{\eta}$ and $\eta \in \{\mu, \nu, \rho\}$.  
A focused axiom is  \emph{context-free} if it has one of the left three forms, otherwise, it is \emph{contextual}. 
\end{dfn}

\begin{exam}
The axioms in Preliminaries are all focused. More examples:
\begin{center}
$\neg 1 \Rightarrow \quad \quad  \quad \quad \Rightarrow \neg 0 \quad \quad  \quad \quad p, \neg p \Rightarrow \quad \quad  \quad \quad  \Rightarrow p, \neg p$
\vspace{5pt}
\linebreak 
$p * \neg p \Rightarrow \quad \quad  \quad \quad  \tGamma, \neg \top \Rightarrow \tDelta \quad \quad  \quad \quad \tGamma \Rightarrow \tDelta, \neg \bot$   
\end{center}
The first five are context-free and the other two are contextual. For a non-example, consider the axiom $(p, \neg p, q \Rightarrow \,)$, where $p$ and $q$ are distinct atoms. This axiom is not focused, as the set of the variables of $p$ and $q$ 
are not equal.
\end{exam}

\begin{rem}
If a single- or multi-conclusion semi-analytic rule has no premises, it is in the form $(\phi \Rightarrow )$ or $( \Rightarrow \phi)$, which is a context-free focused axiom. 
\end{rem}

\begin{rem}\label{InstanceConvention}
For brevity, we fix a notation for a generic instance of a single- or multi-conclusion semi-analytic rule or a focused axiom. We assume the substitution involved in the instance is called $\sigma$ and its application on the contexts and formulas involved are:
$\sigma(\tGamma_i)=\Gamma_i$, $\sigma(\tPi_j)=\Pi_j$, $\sigma(\tDelta_i)=\Delta_i$, $\sigma(\bar{\mu}_{ir})=\bar{\phi}_{ir}$, $\sigma(\bar{\nu}_{ir})=\bar{\psi}_{ir}$, $\sigma(\bar{\nu}_{js})=\bar{\psi}_{js}$, $\sigma(\bar{\rho}_{js})=\bar{\theta}_{js}$, and  $\sigma(\mu)=\phi$. 
\end{rem}

\begin{dfn} \label{DfnStrong}
Let $\mathcal{L} \in \{\mathcal{L}_I^b, \mathcal{L}_I^{u}\}$ and $G$ be a single-conclusion (resp. multi-conclusion) sequent calculus over $\mathcal{L}$. The calculus $G$ is called  \emph{strong}:
  \begin{itemize}
      \item 
      if $\mathcal{L}=\mathcal{L}_I^b$, then $G$ extends $\mathbf{IMALL}$ (resp. $\mathbf{MALL}$), and
      \item
     if $\mathcal{L}=\mathcal{L}_I^{u}$, then $G$ extends $\mathbf{FL_e}$ (resp. $\mathbf{CFL_e}$).
  \end{itemize}
\end{dfn}

\begin{rem} \label{RemStrong}
By Definition \ref{CalculusExtension},
if $G$ is a strong single-conclusion calculus over $\mathcal{L}_I^{u}$, each instance of any axiom of $\mathbf{FL_e}$ is provable in $G$ and each rule of $\mathbf{FL_e}$ is admissible in $G$. Hence, we can augment $G$ with the axioms and rules of $\mathbf{FL_e}$, preserving the logic of $G$. Similarly for the other cases.
\end{rem}

\begin{dfn}
A calculus $G$ over  $\mathcal{L} \in \{\mathcal{L}_I^b, \mathcal{L}_I^{u}\}$ is \emph{(single-conclusion) multi-conclusion semi-analytic} if 
each rule in $G$ is either (single-conclusion) multi-conclusion semi-analytic or $\ruleK$ or $\ruleD$, and its axioms are (single-conclusion  versions of) focused axioms if $\mathcal{L}=\mathcal{L}_I^b$, and (single-conclusion versions of) context-free focused axioms if $\mathcal{L}=\mathcal{L}_I^u$.
\end{dfn}

\section{Craig Interpolation Property} \label{Section Craig}

In this section, we establish the main result that the logic of any single- or multi-conclusion semi-analytic calculus has the Craig interpolation property. 
We first generalize the notion of an interpolant from an implication between two formulas to 
sequents. Then, following the Maehara split-interpolation technique we propose two kinds of interpolants, one for single-conclusion and one for multi-conclusion sequents. Over $\mathcal{L}^b_{I}$ we show that in a strong (single-conclusion) multi-conclusion calculus, the focused axioms have the (single-conclusion) multi-conclusion interpolant and the (single-conclusion) multi-conclusion semi-analytic rules and the modal rules $\ruleK$ and $\ruleD$ respect the existence of such (single-conclusion) multi-conclusion interpolants. A similar approach will be employed for $\mathcal{L}^u_{I}$. 

\begin{dfn} \label{DfnCraig}
A logic $L$ has the \emph{Craig Interpolation Property (CIP)} if for any formulas $\phi$ and $\psi$ if $ (\phi \to \psi) \in L$, then there exists a formula $\theta$ such that $( \phi \to \theta) \in L$, $(\theta \to \psi) \in L$ and $V(\theta) \subseteq V(\phi) \cap V(\psi)$.
\end{dfn}

\begin{dfn} \label{DfnMaehara}
Let $\mathcal{A}$ be a set of meta-sequents.
\begin{itemize}
\item
Let $G$ be a single-conclusion calculus. $G$ has \emph{single-conclusion interpolation} ($\mathcal{A}$ has \emph{$G$-single-conclusion interpolation}), if for any sequent $S$ and any partition $\Sigma \cup \Lambda$ of $S^{a}$, if $S$ is provable in $G$ (if $S$ is an instance of a meta-sequent in $\mathcal{A}$),
then there exists a formula $C$, called an \emph{interpolant}, such that $V(C) \subseteq V(\Sigma) \cap V(\Lambda \cup S^s)$ and
\begin{center}
    $G \vdash \Sigma \Rightarrow C \ \ \ \ \text{ and } \ \ \ \ G \vdash \Lambda, C \Rightarrow S^s$.
\end{center}
\item
Let $G$ be a multi-conclusion calculus. $G$ has \emph{multi-conclusion interpolation} ($\mathcal{A}$ has \emph{$G$-multi-conclusion interpolation}), if for any sequent $S$, any partition $\Sigma \cup \Lambda$ of $S^{a}$ and any partition $\Theta \cup \Delta$ of $S^s$, if $G \vdash S$ (if $S$ is an instance of a meta-sequent in $\mathcal{A}$),
then there is a formula $C$, called an \emph{interpolant}, such that $V(C) \subseteq V(\Sigma \cup \Theta) \cap V(\Lambda \cup \Delta)$ and
\begin{center}
    $G \vdash \Sigma \Rightarrow C, \Theta \ \ \ \  \text{ and } \ \ \ \   G \vdash \Lambda, C \Rightarrow \Delta$
\end{center}
\end{itemize}
\end{dfn}
The interpolation property of a calculus implies the CIP of its logic:
\begin{thm}\label{MaeharaImpliesInterpolation} 
Let $G$ be a single-conclusion (multi-conclusion) sequent calculus for the logic $L$. If $G$ has single-conclusion (multi-conclusion) interpolation, then $L$ has the CIP.
\end{thm}
\begin{proof}
Let $\phi \to \psi \in L$. By Definition \ref{CalculusForLogic}, $G \vdash \phi \Rightarrow \psi$. By Definition \ref{DfnMaehara}, there exists $\theta$ such that $G \vdash \phi \Rightarrow \theta $, $G \vdash \theta \Rightarrow \psi$ and $V(\theta) \subseteq V(\phi) \cap V(\psi)$. Again, by Definition \ref{CalculusForLogic}, $\phi \rightarrow \theta \in L$ and  $\theta \rightarrow \psi \in L$, as required.
\end{proof}

\noindent For focused axioms, the existence of the two kinds of interpolants is ensured:
\begin{thm} \label{MaeharaAxioms}
Let $G$ be a strong single-conclusion calculus over $\mathcal{L} \in \{\mathcal{L}_I^b, \mathcal{L}_I^u\}$ and $\mathcal{A}$ be a subset of its axioms only consisting of 
focused axioms.  
\begin{itemize}
\item
If $\mathcal{L}=\mathcal{L}_I^b$, then $\mathcal{A}$ has $G$-single-conclusion interpolation.
\item
If $\mathcal{L}=\mathcal{L}_I^u$ and $\mathcal{A}$ only has context-free axioms, $\mathcal{A}$ has $G$-single-conclusion interpolation.
\end{itemize}
The claims also hold for the multi-conclusion calculi, if we substitute single-conclusion by multi-conclusion, everywhere.
\end{thm}
\begin{proof}
We prove the single-conclusion case.
Let $S$ be an instance of a meta-sequent $\mathcal{S}$ in $\mathcal{A}$.
We investigate each potential form for $\mathcal{S}$ and in each case and for any partition $S^a=\Sigma \cup \Lambda$, we find a formula $C$ such that $G \vdash (\Sigma \Rightarrow C)$, $G \vdash (\Lambda , C \Rightarrow S^s)$ and $V(C) \subseteq V(\Sigma) \cap V(\Lambda \cup S^s)$.
If $C$ is $0, 1, \bot,$ or $\top$,  the variable condition is obviously satisfied. 
Moreover, as $G$ is strong, it extends $\mathbf{FL_e}$ (resp. $\mathbf{IMALL}$) when $\mathcal{L}=\mathcal{L}^u_I$ (resp. $\mathcal{L}=\mathcal{L}^b_I$). Hence, the axioms and rules of the latter are provable and admissible in $G$, respectively. \\
(1) Suppose $\mathcal{S}=(\mu \Rightarrow \mu)$ and $S=(\phi \Rightarrow \phi)$, where $\phi=\sigma(\mu)$. If $\Sigma = \{\phi\}$ and $\Lambda = \{\}$, taking $C$ as $\phi$ works, as $(\phi \Rightarrow \phi) $ is an instance of $\mathcal{S}$. If $\Sigma = \{\}$ and $\Lambda = \{\phi\}$, take $C$ as $1$. Then, $G \vdash (\Rightarrow 1)$ and $G \vdash (1, \phi \Rightarrow \phi)$. The former is an axiom of $\mathbf{FL_e}$ and hence provable in $G$. The latter is the conclusion of an instance of the rule $(1 w)$, which is admissible in $G$, and the fact that $(\phi \Rightarrow \phi)$ is an instance of $\mathcal{S}$. In both cases, $V(C) \subseteq V(\Sigma) \cap V(\Lambda \cup\phi)$.\\
(2)
If $\mathcal{S}=(\Rightarrow \bar{\nu})$ and $S=(\Rightarrow \bar{\psi})$, where $\bar{\psi}=\sigma(\bar{\nu})$, then $\Sigma=\Lambda=\varnothing$. Take $C$ as $1$. Then $(\Rightarrow 1)$ and $(1 \Rightarrow \bar{\psi})$ are provable in $G$. The former is an axiom of $\mathbf{FL_e}$, while the latter is the conclusion of an instance of the rule $(1 w)$ and the fact that $(\Rightarrow \bar{\psi})$ is an instance of $\calS$.\\
(3)
If $\mathcal{S}$ is of the form  $(\bar{\rho} \Rightarrow)$ and $S$ is $(\bar{\theta} \Rightarrow \,)$, where $\bar{\theta}=\sigma(\bar{\rho})$, then:
\begin{itemize}
\item[$(i)$]
If $\Lambda=\bar{\theta}$ , $\Sigma=\{\}$, set $C=1$. Then, $G \vdash (\Sigma \Rightarrow 1)$ and $G \vdash (\Lambda, 1 \Rightarrow)$.
\item[$(ii)$]
If $\Sigma=\bar{\theta}$,  $\Lambda=\{\}$, set $C=0$. Then, $G \vdash (\Sigma \Rightarrow 0)$ and $G \vdash (\Lambda , 0 \Rightarrow)$.
\item[$(iii)$]
If both $\Sigma$ and $\Lambda$ are nonempty, assume $\bar{\theta}= \theta_1, \ldots, \theta_n$ and w.l.o.g. $\Sigma= \theta_1, \ldots, \theta_i$ and $\Lambda= \theta_{i+1}, \ldots, \theta_n$, for an $1 \leq i \leq n-1$.
Set $C=\bigast \Sigma$. Then $G \vdash \Sigma \Rightarrow C$, by applying the rule $(R *)$ for $i-1$ many times on the sequents $\theta_j \Rightarrow \theta_j$, where $1 \leq j \leq i$. Moreover, $G \vdash (\Lambda, C \Rightarrow )$, by the axiom $\mathcal{S}$ itself and applying the rule $(L *)$ for $i-1$ many times. For the variable condition, if $p \in V(C)$, then $p \in V(\Sigma)$. By Definition \ref{Dfn Focused Axioms}, each pair of the elements of $\bar{\rho}$ have the same set of variables. Therefore, the same also holds for $\bar{\theta}$. As both $\Sigma$ and $\Lambda$ are non-empty, $p \in V(\Lambda)$. Hence $p \in V(\Sigma) \cap V(\Lambda)$.
\end{itemize}
(4)
If $\mathcal{S}=(\tGamma, \bar{\mu} \Rightarrow \tDelta)$ and $S=(\Gamma, \bar{\phi} \Rightarrow \Delta)$, where $\bar{\phi}=\sigma(\bar{\mu})$ and $\mathcal{L}=\mathcal{L}^b_I$, then the strongness of $G$ implies that it extends $\mathbf{IMALL}$. Now:
\begin{itemize}
\item[$(i)$]
If $\bar{\phi} \subseteq \Sigma$, take $C=\bot$. Then, $\bot, \Lambda \Rightarrow \Delta$ is an instance of an axiom in $\mathbf{IMALL}$ and hence provable in $G$. Moreover, substituting $\Sigma - \bar{\phi}$ for $\tGamma$ and $\bot$ for $\tDelta$ and applying $\sigma$ on $\bar{\mu}$ in $\mathcal{S}$, we obtain $G \vdash \Sigma \Rightarrow \bot$.
\item[$(ii)$]
If $\bar{\phi} \subseteq \Lambda$, take $C=\top$. Then $\Sigma \Rightarrow \top$ is an instance of an axiom in $\mathbf{IMALL}$, hence provable in $G$. Moreover, substituting 
$(\Lambda - \bar{\phi}) \cup \{\top\}$ for $\tGamma$ and $\Delta$ for $\tDelta$ and applying $\sigma$ on $\bar{\mu}$ in $\mathcal{S}$ we obtain $G \vdash \top, \Lambda \Rightarrow \Delta$. 
\item[$(iii)$]
If both $\bar{\phi} \cap \Sigma$ and $\bar{\phi} \cap \Lambda$ are nonempty, take $C=\bigast (\Sigma \cap \bar{\phi}) * \top^n$, where $n$ is the cardinality of $\Sigma - \bar{\phi}$. If $\Sigma - \bar{\phi}=\varnothing$, then $n=0$ and by $\top^0$ we mean $1$. First, we prove $G \vdash \Sigma \Rightarrow C$. As $G \vdash (\phi_i \Rightarrow \phi_i)$ for any $\phi_i \in \Sigma \cap \bar{\phi}$, we have $G \vdash \Sigma \cap \bar{\phi} \Rightarrow \bigast (\Sigma \cap \bar{\phi})$. If $\Sigma - \bar{\phi}=\varnothing$, by the axiom $(\, \Rightarrow 1)$ and the rule $(R *)$, we get $G \vdash \Sigma \Rightarrow C$. If $\Sigma - \bar{\phi} \neq \varnothing$, as $G \vdash \Sigma \cap \bar{\phi} \Rightarrow \bigast (\Sigma \cap \bar{\phi})$ and
$G \vdash (\psi \Rightarrow \top)$ for any $\psi \in \Sigma -  \bar{\phi}$,  we can repeatedly apply $(R *)$ to reach $\Sigma \Rightarrow C$.
Second, $G \vdash \Lambda, C \Rightarrow \Delta$. Because the part of $\bar{\phi}$ in $\Sigma$ (and now in $C$) together with the part of $\bar{\phi}$ in $\Lambda$ completes $\bar{\phi}$. Indeed, if we substitute $(\Lambda - \bar{\phi}) \cup \{\top^n\}$ and $\Delta$ for $\tGamma$ and $\tDelta$, respectively and apply $\sigma$ on $\bar{\mu}$ in $\mathcal{S}$, we get $(\Lambda - \bar{\phi}), \top^n, \bar{\phi} \Rightarrow \Delta$ which is
$(\Lambda - \bar{\phi}), \top^n, (\Lambda \cap \bar{\phi}), (\Sigma \cap \bar{\phi}) \Rightarrow \Delta$, by the fact that $\bar{\phi}$ is split between $\Sigma$ and $\Lambda$. Applying $(L *)$ on 
$(\Lambda - \bar{\phi}), \top^n, (\Lambda \cap \bar{\phi}), (\Sigma \cap \bar{\phi}) \Rightarrow \Delta$
repeatedly, one obtains $G \vdash \Lambda, C \Rightarrow \Delta$. For the variables, if $p \in V(C)$ then $p \in V(\Sigma \cap \bar{\phi})$. As $\Sigma \cap \bar{\phi}\neq \emptyset$, $p$ occurs in one of $\bar{\phi}$'s in $\Sigma$. Since there is at least one of $\bar{\phi}$'s in $\Lambda$ and each pair of the elements of $\bar{\mu}$ and hence $\bar{\phi}$ have the same variables, $p \in V(\Lambda)$.
\end{itemize}
(5)
If $\calS=(\tGamma \Rightarrow \bar{\mu} \,)$ and $S=(\Gamma \Rightarrow \bar{\phi})$, where $\bar{\phi}=\sigma(\bar{\mu})$ and $\mathcal{L}=\mathcal{L}^b_I$, then the strongness of $G$ implies that it extends $\mathbf{IMALL}$. Take $C=\top$. Both $\Sigma \Rightarrow \top$ and $\top, \Lambda \Rightarrow \bar{\phi}$ are provable in $G$, where the latter is an instance of the axiom $\mathcal{S}$, substituting the multiset $\{\top\} \cup \Lambda$ for $\tGamma$ and applying $\sigma$ on $\bar{\mu}$.
\end{proof}

\begin{thm} \label{MaeharaSubClassic} 
Let $G= \mathcal{A} \cup H$ be a strong single-conclusion calculus over $\mathcal{L}\in \{\mathcal{L}_I^u, \mathcal{L}_I^b\}$, $\mathcal{A}$ be a set of meta-sequents and $H$ be a single-conclusion semi-analytic calculus.
Then, if $\mathcal{A}$ has $G$-single-conclusion interpolation, $G$ has single-conclusion interpolation.
\end{thm}

\begin{proof} 
As $G$ is strong, in both cases $\mathcal{L}=\mathcal{L}_I^u$ or $\mathcal{L}=\mathcal{L}_I^b$, it extends $\mathbf{FL_e}$ and hence its axioms and rules are provable and admissible in $G$, respectively.
Assume $\mathcal{A}$ has $G$-single-conclusion interpolation. For any provable sequent $S$ in $G$ with the proof $\pi$ and any partition $\Sigma \cup \Lambda$ of $S^{a}$, we find a formula $C$ such that $G \vdash \Sigma \Rightarrow C$, $G \vdash \Lambda, C \Rightarrow S^s$ and $V(C) \subseteq V(\Sigma) \cap V(\Lambda \cup S^s)$.
To prove this, we use an induction on the structure of $\pi$. If $\pi$ is an instance of an axiom $\calS$, either $\calS \in \mathcal{A}$ or it is focused (resp. context-free focused) when $\mathcal{L}=\mathcal{L}^b_I$ (resp. $\mathcal{L}=\mathcal{L}^u_I$). In the former case, use the assumption that $\mathcal{A}$ has $G$-single-conclusion interpolation and in the latter, use Theorem \ref{MaeharaAxioms} to find $C$.
If $\pi$ is not an instance of an axiom, there are three main cases:

\item[$(i)$] The proof $\pi$ ends by applying a left single-conclusion semi-analytic rule:
\footnotesize\begin{center}
 \begin{tabular}{c c}
 \AxiomC{$\{\tPi_j , \bar{\nu}_{js} \Rightarrow \bar{\rho}_{js} \mid 1 \leq j \leq m, 1 \leq s \leq l_j \}$} 
 \AxiomC{$\{\tGamma_i , \bar{\mu}_{ir} \Rightarrow \tDelta_i \mid 1 \leq i \leq n, 1 \leq r \leq k_i \}$}
 \BinaryInfC{$\tPi_1, \dots, \tPi_m, \tGamma_1, \dots, \tGamma_n, \mu \Rightarrow \tDelta_1, \dots, \tDelta_n $}
 \DisplayProof  $(\dagger)$
\end{tabular}
\end{center}
\normalsize
Following the convention in Remark \ref{InstanceConvention}, the instance of the conclusion is in the form $S=(\Pi_1, \dots, \Pi_m, \Gamma_1, \dots, \Gamma_n, \phi \Rightarrow \Delta_1, \dots, \Delta_n)$. For brevity, we omit the domains of the variables $i, j, r, s$, when there is no ambiguity. Denote the multiset $\Gamma_1, \ldots, \Gamma_n$ by $\Gamma$, the multiset $\Pi_1, \ldots, \Pi_m$ by $\Pi$ and $\Delta_1, \ldots, \Delta_n$ by $\Delta$. Let $\Gamma=\Gamma' \cup \Gamma''$ and $\Pi=\Pi' \cup \Pi''$ be the partitions induced by the partition of $S^a$. Set $\Gamma'_i$ as $\Gamma' \cap \Gamma_i$. A similar convention applies to $\Pi$ and $\Delta$.
Now, there are two cases to investigate, either $\phi \in \Lambda$ or $\phi \in \Sigma$:
\item[$\circ$]
If $\phi \in \Lambda$, then $\Sigma =\Gamma' \cup \Pi'$ and $\Lambda =\Gamma'' \cup \Pi''  \cup \{\phi\}$.
We find $C$ that $G \vdash \Gamma' , \Pi'  \Rightarrow C$ and $G \vdash \Gamma'' ,  \Pi'' , \phi , C \Rightarrow \Delta$. The last rule in $\pi$ is of the form
\begin{center}
	  	\AxiomC{$ \{\Pi'_j, \Pi''_j , \bar{\psi}_{js} \Rightarrow \bar{\theta}_{js} \}_{s,j}$}
	  	\AxiomC{$\{ \Gamma'_i, \Gamma''_i , \bar{\phi}_{ir} \Rightarrow \Delta_i \}_{r,i}$}
  		\BinaryInfC{$\Pi' , \Pi'' ,   \Gamma' , \Gamma'' , \phi \Rightarrow \Delta$} 
  		\DisplayProof
\end{center}
By the induction hypothesis, there are $C_{js}$ and $D_{ir}$ for each $i, j, r,s$ such that
\begin{center}
	  	$G \vdash \Pi'_j \Rightarrow C_{js} $
\quad \quad \quad \quad 
	  	$G \vdash  \Pi''_j , \bar{\psi}_{js}, C_{js} \Rightarrow \bar{\theta}_{js}$
\end{center}
\begin{center}
  	\begin{tabular}{c}
	  $G \vdash  \Gamma'_i \Rightarrow D_{ir} $
\quad \quad \quad \quad
	  	$G \vdash  \Gamma''_i, \bar{\phi}_{ir}, D_{ir} \Rightarrow \Delta_i$
	  	\end{tabular}
\end{center}
and $V(C_{js}) \subseteq V(\Pi'_j) \cap V(\Pi''_j \cup \bar{\psi}_{js} \cup \bar{\theta}_{js})$
and $V(D_{ir}) \subseteq V(\Gamma'_i) \cap V(\Gamma''_i \cup \bar{\phi}_{ir} \cup \Delta_i)$.
Applying $(R \wedge)$ on the left sequents and $(L \wedge)$ on the right ones, we get for each $i,j,r,s$:
\begin{center}
	  	$G \vdash  \Pi'_j \Rightarrow  \bigwedge \limits_s C_{js} $ \ (1)
\quad \quad \quad \quad 
	  	$G \vdash  \Pi''_j , \bar{\psi}_{js}, \bigwedge \limits_s  C_{js} \Rightarrow \bar{\theta}_{js}$ \ (3)
\end{center}
\begin{center}
	  $G \vdash  \Gamma'_i \Rightarrow \bigwedge \limits_r D_{ir} $ \ (2)
\quad \quad \quad \quad 
	  	$G \vdash \Gamma''_i , \bar{\phi}_{ir}, \bigwedge \limits_r  D_{ir} \Rightarrow \Delta_i$ \ (4)
\end{center}
Applying the rule $(R *)$ on the sequents (1) and (2), we obtain
\begin{center}
$G \vdash  \Pi'_1, \ldots, \Pi'_m, \Gamma'_1, \ldots, \Gamma'_n \Rightarrow (\bigast \limits_j \bigwedge \limits_s C_{js}) * (\bigast \limits_i \bigwedge \limits_r  D_{ir})$.
\end{center}
Apply the rule $(\dagger)$ on the sequents (3) and (4), where $\tilde{\Pi}_j$ is substituted by $\Pi_j^{\prime \prime} \cup\{\bigwedge_s C_{j s}\}$ and $\tGamma_i$ is substituted by $\Gamma_i^{\prime \prime} \cup\{\bigwedge_r D_{ir}\}$.
This is doable, as the contexts are free for any substitution and $\bigwedge_s C_{js}$ (resp. $\bigwedge_r D_{ir}$) depends only on $j$ and not on $s$ (resp. only on $i$ and not on $r$). Then, apply $(L*)$:
\begin{center}
$ G \vdash  \Pi''_1, \ldots, \Pi''_m, \Gamma''_1, \ldots, \Gamma''_n , (\bigast \limits_j \bigwedge \limits_s C_{js}) * (\bigast \limits_i \bigwedge \limits_r  D_{ir}) , \phi \Rightarrow \Delta_1, \ldots, \Delta_n$.
\end{center}
Define $C$ as $(\bigast \limits_j \bigwedge \limits_s C_{js}) * (\bigast \limits_i \bigwedge \limits_r  D_{ir})$. Clearly, $C$ satisfies the provability conditions of the interpolation. To show $V(C) \subseteq V(\Sigma) \cap V(\Lambda \cup \Delta)$, an atom $p$ is in $C$ if{f} it is in one of $C_{js}$'s or $D_{ir}$'s. If it is in $C_{js}$, by the induction hypothesis, it is both in $\Pi'_j$ (and hence in $\Sigma$) and $\Pi''_j \cup \bar{\psi}_{js} \cup \bar{\theta}_{js}$. If it is in $\Pi''_j$, then it is in $\Lambda$ and if it is in either $\bar{\psi}_{js}$ or $\bar{\theta}_{js}$, since the rule is occurrence-preserving, it also appears in $\phi$ implying it appears in $\Lambda$. The case $p \in D_{ir}$ is similar.

\item[$\circ$]
If $\phi \in \Sigma$, then $\Sigma =\Gamma' \cup \Pi' \cup \{\phi\}$ and $\Lambda =\Gamma''\cup \Pi''$. We find $C$ that $G \vdash \Gamma' , \Pi' , \phi  \Rightarrow C$ and $G \vdash \Gamma'' ,  \Pi'' ,  C \Rightarrow \Delta$. As $S$ is  single-conclusion and $\Delta=\Delta_1, \ldots, \Delta_n$, at most one of $\Delta_i$'s can be nonempty. W.l.o.g., suppose $\Delta_1 = \Delta$ and for $i \neq 1$, we have $\Delta_i = \emptyset$.
Thus, the last rule in $\pi$ is:
\begin{center}
  	\begin{tabular}{c}
	  	\AxiomC{$\{ \Pi'_j, \Pi''_j , \bar{\psi}_{js} \Rightarrow \bar{\theta}_{js} \}_{s,j}$}
	  	\AxiomC{$\{ \Gamma'_i, \Gamma''_i , \bar{\phi}_{ir} \Rightarrow  \}_{r,i \neq 1}$}
	  	\AxiomC{$\{ \Gamma'_1, \Gamma''_1 , \bar{\phi}_{1r} \Rightarrow \Delta \}_r $}
  		\TrinaryInfC{$\Pi' , \Pi'' ,   \Gamma' , \Gamma'' , \phi \Rightarrow \Delta$}
  		\DisplayProof
\end{tabular}
\end{center}
By the induction hypothesis, there are formulas $C_{js}$ and $D_{ir}$ for each $i \neq 1$ and $j, s, r$ and $D_{1r}$ for each $r$ such that the following are provable in $G$:
\begin{center}  	
	  	$\Pi'_j , \bar{\psi}_{js}, C_{js} \Rightarrow \bar{\theta}_{js}$
\quad \quad \quad \quad 
    $\Gamma'_i, \bar{\phi}_{ir}, D_{ir} \Rightarrow $	 
\quad \quad \quad \quad
	 $\Gamma'_1, \bar{\phi}_{1r} \Rightarrow D_{1r}$ 	
\end{center}
\begin{center}  
  	$\Pi''_j \Rightarrow C_{js} $
\quad \quad \quad \quad
	  	 $ \Gamma''_i \Rightarrow D_{ir} $	
\quad \quad \quad \quad
	  	$ \Gamma''_1 , D_{1r} \Rightarrow \Delta$
\end{center}
and $V(C_{js}) \subseteq V(\Pi'_j \cup \bar{\psi}_{js} \cup \bar{\theta}_{js}) \cap V(\Pi''_j)$, $V(D_{ir}) \subseteq V(\Gamma'_i \cup \bar{\phi}_{ir}) \cap V(\Gamma''_i)$, for $i \neq 1$ 
and $V(D_{1r}) \subseteq V(\Gamma'_1  \cup \bar{\phi}_{1r}) \cap V(\Gamma''_1 \cup 
\Delta)$.
Applying $(L \wedge)$, $(R \wedge)$, $(R \vee)$ and $(L \vee)$, all admissible in $G$, for any $i \neq 1$, the following are provable in $G$:
\begin{center}
	  	$\Pi'_j , \bar{\psi}_{js}, \bigwedge \limits_s C_{js} \Rightarrow \bar{\theta}_{js}$ \ (5)
\ \ \ \ 
    $\Gamma'_i, \bar{\phi}_{ir}, \bigwedge \limits_r D_{ir} \Rightarrow $ \ (6)
\ \ \ \ 
    $\Gamma'_1, \bar{\phi}_{1r} \Rightarrow \bigvee \limits_r D_{1r}$  \ (7)
\end{center}
\begin{center}
  	$\Pi''_j \Rightarrow \bigwedge \limits_s C_{js} $ \ (8)
\ \ \ \ 
	  	 $ \Gamma''_i \Rightarrow \bigwedge \limits_r D_{ir} $ \ (9)	  	 
\ \ \ \ 
	  	$ \Gamma''_1 , \bigvee \limits_r D_{1r} \Rightarrow  \Delta$ \ (10)
\end{center}
By substituting (5), (6), and (7) in the rule $(\dagger)$, we get 
\begin{center}
$G \vdash  \Pi', \Gamma', \bigcup_{j} \{\bigwedge \limits_s C_{js}\}, \bigcup_{i \neq 1} \{\bigwedge \limits_r D_{ir}\}, \phi \Rightarrow \bigvee \limits_r D_{1r}. $
\end{center}
Applying the rule $(L *)$ and then $(R \to)$, we get
\begin{center}
$ G \vdash  \Pi', \Gamma', \phi \Rightarrow (\bigast \limits_{i \neq 1} \bigwedge \limits_r D_{ir} ) * (\bigast \limits_j \bigwedge \limits_s C_{js} ) \to \bigvee \limits_r D_{1r}$.
\end{center}
Now, by the rules $(R *)$ and $(L \to)$ on the sequents (8), (9), and (10), we get
\begin{center}
$ G \vdash  \Pi'' , \Gamma'' , (\bigast \limits_{i \neq 1} \bigwedge \limits_r D_{ir} ) * (\bigast \limits_j \bigwedge \limits_s C_{js} ) \to \bigvee \limits_r D_{1r} \Rightarrow \Delta.$
\end{center}
Define $C=(\bigast \limits_{i \neq 1} \bigwedge \limits_r D_{ir} ) * (\bigast \limits_j \bigwedge \limits_s C_{js} ) \to \bigvee \limits_r D_{1r}$. Clearly, $C$ satisfies the provability conditions. To prove $V(C) \subseteq V(\Sigma) \cap V(\Lambda \cup \Delta)$, an atom $p$ is in $C$ if and only if it is either in one of $C_{js}$'s or $D_{ir}$'s (for $i \neq 1$) or in $D_{1r}$'s. By the induction hypothesis, if $p \in C_{js}$, it is both in $\Pi'_j \cup \bar{\psi}_{js} \cup \bar{\theta}_{js}$ and $\Pi''_j$ (hence $p \in \Lambda \cup \Delta$). If $p \in \Pi'_j$, then $p \in \Sigma$. If $p \in \bar{\psi}_{js} \cup \bar{\theta}_{js}$, by the occurrence-preservation, $p \in \phi$, hence $p \in \Sigma$. Thus, $p$ is in both $\Sigma=\Gamma' \cup \Pi' \cup \{\phi\}$ and $\Lambda \cup \Delta=\Gamma'' \cup \Pi'' \cup \Delta$. The case where $p$ is in $D_{ir}$ (for $i \neq 1$) or $D_{1r}$ is similar.

\item[$(ii)$]
The proof $\pi$ ends by applying a right single-conclusion semi-analytic rule:
\small\begin{center}
 \begin{tabular}{c }
 \AxiomC{$\{\tGamma_i , \bar{\mu}_{ir} \Rightarrow \bar{\nu}_{ir} \mid 1 \leq i \leq n, 1 \leq r \leq k_i\}$}
 \UnaryInfC{$\tGamma_1, \dots, \tGamma_n \Rightarrow \mu $}
 \DisplayProof $(\ddagger)$
\end{tabular}
\end{center}
\normalsize
Following the convention in Remark \ref{InstanceConvention}, the conclusion of the instance is in the form $S= (\Gamma_1, \ldots, \Gamma_n \Rightarrow \phi)$. Let $\Sigma_i=\Gamma_i \cap \Sigma$ and $\Lambda_i=\Gamma_i \cap \Lambda$, for any $i$. We find $C$ satisfying $G \vdash \Sigma \Rightarrow C$ and $G \vdash \Lambda , C \Rightarrow \phi$. The last rule of $\pi$ is:
\begin{center}
  	\begin{tabular}{c}
	  	\AxiomC{$\{ \Sigma_i, \Lambda_i , \bar{\phi}_{ir} \Rightarrow \bar{\psi}_{ir} \}_{r,i}$}
  		\UnaryInfC{$ \Sigma , \Lambda  \Rightarrow \phi$}
  		\DisplayProof 
\end{tabular}
\end{center}
By the induction hypothesis, there are formulas $C_{ir}$ that for each $i$ and $r$:
\begin{center}
  $G \vdash \Lambda_i , C_{ir} , \bar{\phi}_{ir} \Rightarrow \bar{\psi}_{ir}$
\quad \quad \quad \quad 
   $G \vdash  \Sigma_i \Rightarrow C_{ir}$.
\end{center}
Applying the rules $(L \wedge)$ and $(R \wedge)$, we have
\begin{center}
  $G \vdash  \Lambda_i ,\bigwedge \limits_r  C_{ir} , \bar{\phi}_{ir} \Rightarrow \bar{\psi}_{ir}$
\quad \quad \quad \quad 
   $G \vdash  \Sigma_i \Rightarrow \bigwedge \limits_r C_{ir}$.
\end{center}
Substituting the left sequents for each $i,r$ in the rule $(\ddagger)$ and then applying $(L *)$, we obtain 
$G \vdash  \Lambda , \bigast \limits_i (\bigwedge \limits_r  C_{ir}) \Rightarrow \phi.$
Moreover, applying $(R *)$ on the sequents $\Sigma_i \Rightarrow \bigwedge \limits_r C_{ir}$, for $1 \leq i \leq n$, we get $G \vdash \Sigma \Rightarrow \bigast \limits_i (\bigwedge \limits_r C_{ir})$. Hence, taking $C=\bigast \limits_i (\bigwedge \limits_r C_{ir})$, the provability conditions of the interpolant are satisfied. Checking the variable condition is easy and similar to the previous cases.

\item[$(iii)$]
The proof $\pi$ ends with applying either the rule $\ruleK$ or $\ruleD$:
\begin{center}
  	\begin{tabular}{c}
	  	\AxiomC{$\Gamma_\alpha,  \Gamma_{\beta_1}, \dots,  \Gamma_{\beta_k}, !^{\gamma_1} \Pi_{\gamma_1}, \dots, !^{\gamma_m} \Pi_{\gamma_m} \Rightarrow \Delta $} 
  		\UnaryInfC{$ !^\alpha \Gamma_\alpha, !^{\beta_1} \Gamma_{\beta_1}, \dots, !^{\beta_k} \Gamma_{\beta_k}, !^{\gamma_1} \Pi_{\gamma_1}, \dots, !^{\gamma_m} \Pi_{\gamma_m} \Rightarrow !^\alpha \Delta $}
  		\DisplayProof 
\end{tabular}
\end{center}
\normalsize where $\Delta$ is either one formula (in the case of $\ruleK$) or $|\Delta| \leq 1$ (in the case of $\ruleD$). Let $\Sigma=!^\alpha \Gamma'_\alpha, !^{\beta_1} \Gamma'_{\beta_1}, \dots, !^{\beta_k} \Gamma'_{\beta_k}, !^{\gamma_1} \Pi'_{\gamma_1}, \dots, !^{\gamma_m} \Pi'_{\gamma_m}$ and $\Lambda=!^\alpha \Gamma''_\alpha, !^{\beta_1} \Gamma''_{\beta_1}, \dots, !^{\beta_k} \Gamma''_{\beta_k}, !^{\gamma_1} \Pi''_{\gamma_1}, \dots, !^{\gamma_m} \Pi''_{\gamma_m}$, where the partitions $\Gamma_{\delta}=\Gamma'_{\delta} \cup \Gamma''_{\delta}$ and $\Pi_{\delta}=\Pi'_{\delta} \cup \Pi''_{\delta}$ are induced by the partition of $S^a$ into $\Sigma$ and $\Lambda$, for any $\delta \in \{\alpha, \beta_1, \ldots, \beta_k, \gamma_1, \ldots, \gamma_m\}$. We find a formula $C$ satisfying
\[
G \vdash !^\alpha \Gamma'_\alpha,  !^{\beta_1} \Gamma'_{\beta_1}, \dots, !^{\beta_k} \Gamma'_{\beta_k}, !^{\gamma_1} \Pi'_{\gamma_1}, \dots, !^{\gamma_m} \Pi'_{\gamma_m} \Rightarrow C, 
\]
\[
G \vdash C, !^\alpha \Gamma''_\alpha,  !^{\beta_1} \Gamma''_{\beta_1}, \dots, !^{\beta_k} \Gamma''_{\beta_k}, !^{\gamma_1} \Pi''_{\gamma_1}, \dots, !^{\gamma_m} \Pi''_{\gamma_m} \Rightarrow !^\alpha \Delta .
\]
By the induction hypothesis, there is a formula $E$ such that
\[
G \vdash  \Gamma'_\alpha,  \Gamma'_{\beta_1}, \dots, \Gamma'_{\beta_k}, !^{\gamma_1} \Pi'_{\gamma_1}, \dots, !^{\gamma_m} \Pi'_{\gamma_m} \Rightarrow E, 
\]
\[
G \vdash E, \Gamma''_\alpha,   \Gamma''_{\beta_1}, \dots, \Gamma''_{\beta_k}, !^{\gamma_1} \Pi''_{\gamma_1}, \dots, !^{\gamma_m} \Pi''_{\gamma_m} \Rightarrow  \Delta .
\]
If the last rule used in $\pi$ is an instance of $\ruleK$ (resp. $\ruleD$), apply $\ruleK$ (resp. $\ruleD$) on both of the above sequents to get
\[
G \vdash !^\alpha \Gamma'_\alpha,  !^{\beta_1} \Gamma'_{\beta_1}, \dots, !^{\beta_k} \Gamma'_{\beta_k}, !^{\gamma_1} \Pi'_{\gamma_1}, \dots, !^{\gamma_m} \Pi'_{\gamma_m} \Rightarrow  !^\alpha E, 
\]
\[
G \vdash  !^\alpha E, !^\alpha \Gamma''_\alpha,  !^{\beta_1} \Gamma''_{\beta_1}, \dots, !^{\beta_k} \Gamma''_{\beta_k}, !^{\gamma_1} \Pi''_{\gamma_1}, \dots, !^{\gamma_m} \Pi''_{\gamma_m} \Rightarrow !^\alpha \Delta .
\]
It is easy to see that the formula $C= !^\alpha E$ serves as the interpolant. 
\end{proof}

The next theorem proves the multi-conclusion version of Theorem \ref{MaeharaSubClassic}.

\begin{thm} \label{MaeharaClassic}
Let $G= \mathcal{A} \cup H$ be a strong multi-conclusion calculus over $\mathcal{L}\in \{\mathcal{L}_I^u, \mathcal{L}_I^b\}$, $\mathcal{A}$ a set of meta-sequents and $H$ a multi-conclusion semi-analytic calculus.
Then, if $\mathcal{A}$ has $G$-multi-conclusion interpolation, $G$ has multi-conclusion interpolation.
\end{thm}

\begin{proof}
As $G$ is strong, in both cases  $\mathcal{L}=\mathcal{L}_I^u$ or $\mathcal{L}=\mathcal{L}_I^b$, it extends $\mathbf{CFL_e}$ and hence its axioms and rules are provable and admissible in $G$, respectively. In particular, the rules $(L+)$ and $(R+)$, provable in $\mathbf{CFL_e}$, are admissible in $G$. Assume $\mathcal{A}$ has $G$-multi-conclusion interpolation. For any provable sequent $S$ in $G$ with the proof $\pi$ and any partitions $S^{a}=\Sigma \cup \Lambda$ and $S^{s}=\Theta \cup \Omega$, we find a formula $C$ such that $G \vdash \Sigma \Rightarrow C, \Theta$, $G \vdash \Lambda, C \Rightarrow \Omega$ and $V(C) \subseteq V(\Sigma \cup \Theta) \cap V(\Lambda \cup \Omega)$. 
To find $C$, we use an induction on the structure of $\pi$. If $\pi$ is an instance of an axiom, we proceed as in the proof of Theorem \ref{MaeharaSubClassic}. Otherwise, based on the last rule in $\pi$, there are three main cases:

\item[$(i)$]
As the case for the right rules is similar, we only investigate the case where the proof $\pi$ ends by applying a left multi-conclusion semi-analytic rule:

\small \begin{center}
 \begin{tabular}{c}
 \AxiomC{$\{ \tGamma_i , \bar{\mu}_{ir} \Rightarrow \bar{\nu}_{ir}, \tDelta_i \mid 1 \leq i \leq n, 1 \leq r \leq k_i \}$}
 \UnaryInfC{$\tGamma_1, \dots, \tGamma_n, \mu \Rightarrow \tDelta_1, \dots, \tDelta_n $}
 \DisplayProof $(\dagger)$
\end{tabular}
\end{center}
\normalsize
Following the convention in Remark \ref{InstanceConvention}, we have $S=(\Gamma_1, \ldots, \Gamma_n, \phi \Rightarrow \Delta_1, \ldots, \Delta_n)$. Let $\Gamma=\Gamma_1, \ldots, \Gamma_n$ and $\Delta=\Delta_1, \ldots, \Delta_n$ and set $\Gamma=\Gamma' \cup \Gamma''$ as the partition induced by the partition $S^a=\Sigma \cup \Lambda$. Set $\Gamma'_i$ and $\Theta_i$ as $\Gamma' \cap \Gamma_i$ and $\Theta \cap \Delta_i$, respectively and use a similar convention to define $\Gamma''_i$ and $\Omega_i$.
\item[$\circ$]
Assume $\phi \in \Sigma$. Hence, $\Sigma=\Gamma' \cup \{\phi\}$ and $\Lambda=\Gamma''$. We find $C$ that $G \vdash \Gamma' ,  \phi  \Rightarrow C, \Theta$ and $G \vdash \Gamma'' ,  C \Rightarrow \Omega$. As the last rule is of the form
\begin{center}
  	\begin{tabular}{c}
	  	\AxiomC{$\{ \Gamma'_i, \Gamma''_i , \bar{\phi}_{ir} \Rightarrow \bar{\psi}_{ir}, \Theta_i, \Omega_i \}_{r,i}$}
  		\UnaryInfC{$ \Gamma' , \Gamma'' , \phi \Rightarrow \Theta , \Omega$} 
  		\DisplayProof 
\end{tabular}
\end{center}
by the induction hypothesis, for each $i$ and $r$, there exists a formula $C_{ir}$ that
\begin{center}
  	$G \vdash \Gamma'_i, \bar{\phi}_{ir} \Rightarrow \bar{\psi}_{ir}, C_{ir}, \Theta_i$	 
	\quad \quad \quad \quad 
	  	 $G \vdash \Gamma''_i, C_{ir} \Rightarrow \Omega_i$
\end{center}
and $V(C_{ir}) \subseteq V(\Gamma'_i \cup \bar{\phi}_{ir} \cup \bar{\psi}_{ir} \cup \Theta_i) \cap V(\Gamma''_i \cup \Omega_i)$. Applying $(R \vee)$ and $(L \vee)$, we have the following sequents provable in $G$, for every $i$ 
\begin{center}
  	$ \Gamma'_i, \bar{\phi}_{ir} \Rightarrow \bar{\psi}_{ir}, \bigvee \limits_r C_{ir}, \Theta_i$	 \ (1)
\quad \quad \quad \quad \quad
	  	 $\Gamma''_i, \bigvee \limits_r C_{ir} \Rightarrow \Omega_i$ \ (2)
\end{center}
If we substitute the sequents in (1), for all $i$'s, in the original rule $(\dagger)$, we get
\begin{center}
$G \vdash \Gamma'_1, \ldots, \Gamma'_n, \phi \Rightarrow \bigvee \limits_r C_{1r} , \ldots, \bigvee \limits_r C_{nr}, \Theta_1, \ldots, \Theta_n  $.
\end{center}
Applying $(R +)$, we get
$G \vdash \Gamma', \phi \Rightarrow \bigplus \limits_i \bigvee \limits_r C_{ir} , \Theta$. Moreover, applying $(L +)$ on the sequents in (2), for all $i$'s, we obtain
$G \vdash \Gamma'', \bigplus \limits_i \bigvee \limits_r C_{ir} \Rightarrow \Omega  $.
Take $C=\bigplus \limits_i \bigvee \limits_r C_{ir}$. It is easy to see that $C$ is the interpolant.

\item[$\circ$]
Assume $\phi \in \Lambda$. Hence, $\Sigma=\Gamma' $ and $\Lambda=\Gamma'' \cup \{\phi\}$.
We find $C$ that $G \vdash \Gamma'  \Rightarrow C, \Theta$ and $G \vdash \Gamma'' ,  \phi , C \Rightarrow \Omega$. The last rule is of the form
\begin{center}
  	\begin{tabular}{c}
	  	\AxiomC{$\{ \Gamma'_i, \Gamma''_i , \bar{\phi}_{ir} \Rightarrow \bar{\psi}_{ir}, \Theta_i, \Omega_i \}_{r,i}$}
  		\UnaryInfC{$  \Gamma' , \Gamma'' , \phi \Rightarrow \Theta, \Omega$}
  		\DisplayProof
\end{tabular}
\end{center}
By the induction hypothesis, for every $i$ and $r$, there is a formula $C_{ir}$ that
\begin{center}
	  $G \vdash \Gamma'_i \Rightarrow C_{ir}, \Theta_i $
\quad \quad \quad \quad 
	  	$G \vdash\Gamma''_i, \bar{\phi}_{ir}, C_{ir} \Rightarrow \bar{\psi}_{ir}, \Omega_i$
\end{center}
and $V(C_{ir}) \subseteq V(\Gamma'_i \cup \Theta_i) \cap V(\Gamma''_i \cup \bar{\phi}_{ir} \cup \bar{\psi}_{ir} \cup \Omega_i)$. Applying the rules $(R \wedge)$ and $(L \wedge)$, the following are provable in $G$, for every $i$:
\begin{center}
	  $ \Gamma'_i \Rightarrow \bigwedge \limits_r C_{ir}, \Theta_i $ \ (3)
	  	\quad \quad \quad \quad 
	  	$\Gamma''_i, \bar{\phi}_{ir}, \bigwedge \limits_r C_{ir} \Rightarrow \bar{\psi}_{ir}, \Omega_i$ \ (4)
\end{center}
Apply $(R *)$ on the sequents in (3) for all $i$'s to get
$ G \vdash \Gamma' \Rightarrow  \bigast \limits_i \bigwedge \limits_r C_{ir}, \Theta $. Now, substitute the sequents in (4) in the rule $(\dagger)$, and then apply $(L *)$ to get
$G \vdash \Gamma'', \phi , \bigast \limits_i \bigwedge \limits_r C_{ir} \Rightarrow \Omega$.
Take $C=\bigast \limits_i \bigwedge \limits_r C_{ir}$ as the interpolant. 
\item[$(ii)$]
Suppose the last rule is an instance of the rule $\ruleK$ in the form
\begin{center}
  	\begin{tabular}{c}
	  	\AxiomC{$\Gamma_\alpha,  \Gamma_{\beta_1}, \dots,  \Gamma_{\beta_k}, !^{\gamma_1} \Pi_{\gamma_1}, \dots, !^{\gamma_m} \Pi_{\gamma_m} \Rightarrow \phi $} 
  		\UnaryInfC{$ !^\alpha \Gamma_\alpha, !^{\beta_1} \Gamma_{\beta_1}, \dots, !^{\beta_k} \Gamma_{\beta_k}, !^{\gamma_1} \Pi_{\gamma_1}, \dots, !^{\gamma_m} \Pi_{\gamma_m} \Rightarrow !^\alpha \phi $}
  		\DisplayProof 
\end{tabular}
\end{center}
\normalsize Let $\Sigma=!^\alpha \Gamma'_\alpha, !^{\beta_1} \Gamma'_{\beta_1}, \dots, !^{\beta_k} \Gamma'_{\beta_k}, !^{\gamma_1} \Pi'_{\gamma_1}, \dots, !^{\gamma_m} \Pi'_{\gamma_m}$ and set $\Lambda$ as the multiset $!^\alpha \Gamma''_\alpha, !^{\beta_1} \Gamma''_{\beta_1}, \dots, !^{\beta_k} \Gamma''_{\beta_k}, !^{\gamma_1} \Pi''_{\gamma_1}, \dots, !^{\gamma_m} \Pi''_{\gamma_m}$, where the partitions $\Gamma_{\delta}=\Gamma'_{\delta} \cup \Gamma''_{\delta}$ and $\Pi_{\delta}=\Pi'_{\delta} \cup \Pi''_{\delta}$ are induced by the partition of $S^a$ into $\Sigma$ and $\Lambda$, for any $\delta \in \{\alpha, \beta_1, \ldots, \beta_k, \gamma_1, \ldots, \gamma_m\}$. Now, either $\Omega= \{!^\alpha \phi\}$ and $\Theta= \emptyset$, or $\Theta= \{!^\alpha \phi\}$ and $\Omega= \emptyset$. The former case is similar to the case $(iii)$ in the proof of Theorem \ref{MaeharaSubClassic}. For the latter case, we find $C$ such that
\[
G \vdash !^\alpha \Gamma'_\alpha,  !^{\beta_1} \Gamma'_{\beta_1}, \dots, !^{\beta_k} \Gamma'_{\beta_k}, !^{\gamma_1} \Pi'_{\gamma_1}, \dots, !^{\gamma_m} \Pi'_{\gamma_m} \Rightarrow C, !^\alpha \phi
\]
\[
G \vdash C, !^\alpha \Gamma''_\alpha,  !^{\beta_1} \Gamma''_{\beta_1}, \dots, !^{\beta_k} \Gamma''_{\beta_k}, !^{\gamma_1} \Pi''_{\gamma_1}, \dots, !^{\gamma_m} \Pi''_{\gamma_m} \Rightarrow 
\]
By the induction hypothesis, there is a formula $E$ such that 
\[
G \vdash  \Gamma'_\alpha,  \Gamma'_{\beta_1}, \dots, \Gamma'_{\beta_k}, !^{\gamma_1} \Pi'_{\gamma_1}, \dots, !^{\gamma_m} \Pi'_{\gamma_m} \Rightarrow E, \phi
\]
\[
G \vdash E, \Gamma''_\alpha,   \Gamma''_{\beta_1}, \dots, \Gamma''_{\beta_k}, !^{\gamma_1} \Pi''_{\gamma_1}, \dots, !^{\gamma_m} \Pi''_{\gamma_m} \Rightarrow  
\]
Now, consider the following proof tree in $G$:
\begin{center}
\AxiomC{$\Gamma'_\alpha,  \Gamma'_{\beta_1}, \dots, \Gamma'_{\beta_k}, !^{\gamma_1} \Sigma'_{\gamma_1}, \dots, !^{\gamma_m} \Sigma'_{\gamma_m} \Rightarrow E, \phi$ }
\AxiomC{$0 \Rightarrow$}
\RightLabel{\scriptsize{$(L \to)$}}
\BinaryInfC{$\Gamma'_\alpha,  \Gamma'_{\beta_1}, \dots, \Gamma'_{\beta_k}, !^{\gamma_1} \Sigma'_{\gamma_1}, \dots, !^{\gamma_m} \Sigma'_{\gamma_m}, \neg E \Rightarrow \phi$}
\RightLabel{\scriptsize{$(\ruleK)$}}
\UnaryInfC{$!^\alpha \Gamma'_\alpha,  !^{\beta_1}\Gamma'_{\beta_1}, \dots, !^{\beta_k} \Gamma'_{\beta_k}, !^{\gamma_1} \Sigma'_{\gamma_1}, \dots, !^{\gamma_m} \Sigma'_{\gamma_m}, !^{\alpha}\neg E \Rightarrow !^{\alpha}\phi$}
\RightLabel{\scriptsize{$(0w)$}}
\UnaryInfC{$!^\alpha \Gamma'_\alpha,  !^{\beta_1}\Gamma'_{\beta_1}, \dots, !^{\beta_k} \Gamma'_{\beta_k}, !^{\gamma_1} \Sigma'_{\gamma_1}, \dots, !^{\gamma_m} \Sigma'_{\gamma_m}, !^{\alpha}\neg E \Rightarrow !^{\alpha}\phi, 0$}
\RightLabel{\scriptsize{$(R \to)$}}
\UnaryInfC{$!^\alpha \Gamma'_\alpha,  !^{\beta_1}\Gamma'_{\beta_1}, \dots, !^{\beta_k} \Gamma'_{\beta_k}, !^{\gamma_1} \Sigma'_{\gamma_1}, \dots, !^{\gamma_m} \Sigma'_{\gamma_m} \Rightarrow \neg !^\alpha \neg E, !^{\alpha}\phi$}
\DisplayProof
\end{center}
Similarly, we get a proof tree in $G$ for
\[
\neg !^\alpha \neg E, !^\alpha \Gamma''_\alpha,  !^{\beta_1} \Gamma''_{\beta_1}, \dots, !^{\beta_k} \Gamma''_{\beta_k}, !^{\gamma_1} \Sigma''_{\gamma_1}, \dots, !^{\gamma_m} \Sigma''_{\gamma_m} \Rightarrow
\]
Take $C= \neg !^\alpha \neg E$ as the interpolant. The case for $\ruleD$ is similar.
\end{proof}

Combining Theorem \ref{MaeharaSubClassic} with Theorem \ref{MaeharaClassic}, we have:
\begin{thm}\label{MainThm}
Let $\mathcal{L} \in \{\mathcal{L}_I^b, \mathcal{L}_I^u\}$ and $G$ be a strong single-conclusion (multi-conclusion) semi-analytic calculus over $\mathcal{L}$. Then $G$ has single-conclusion  (multi-conclusion) interpolation.  
\end{thm}

To transfer Theorem \ref{MainThm} from the sequent calculi to logics, we first need:
\begin{lem}\label{FLAdmissibility}
Let $L \in \{\mathsf{FL_e} , \mathsf{CFL_e} , \mathsf{IMALL}, \mathsf{MALL}\}$ be a logic, the calculus $G_L \in \{\mathbf{FL_e} , \mathbf{CFL_e} , \mathbf{IMALL}, \mathbf{MALL}\}$ be its corresponding calculus and $G$ be a calculus for a logic $M \supseteq L$. Then, $G$ extends $G_L$.
\end{lem}
\begin{proof}
Observe $M$ is closed under $*$: if $\phi, \psi \in M$, as $\phi \to (\psi \to \phi * \psi) \in L \subseteq M$ and $M$ is closed under modus ponens, $\phi * \psi \in M$. 
To prove the lemma, by Definition \ref{CalculusExtension}, we need to show that each instance of the axioms of $G_L$ and each rule of $G_L$ is provable and admissible in $G$, respectively. If $S$ is an instance of an axiom in $G_L$, by Definition \ref{CalculusForLogic}, $I(S)$ is in $L$ and hence in $M$. Thus, $G \vdash S$. Let $R=$ \AxiomC{$S_1, \ldots, S_n$}
 \UnaryInfC{$S$}
 \DisplayProof
be an instance of a rule in $G_L$. Then:
    \item[$\bullet$] Let $R$ be an instance of the conjunction or disjunction rules. Clearly,
$\bigwedge_{i=1}^n I(S_i) \to I(S)$    
is in $L$, hence in $M$. If $G \vdash S_i$ for all $1 \leq i \leq n$, we get $I(S_i) \in M$ and by Definition \ref{DfnLogic}, $ \bigwedge_{i=1}^n I(S_i) \in M$. As $M$ is closed under modus ponens, $ I(S) \in M$ which implies $G \vdash S$, by Definition \ref{CalculusForLogic}. 

\item[$\bullet$] Let $R$ be an instance of any of the cut, fusion or implication rules or $(0w)$ or $(1w)$. Then,  the formula
$\bigast_{i=1}^n I(S_i) \to I(S)$    
is in $L$ and hence in $M$. If $G \vdash S_i$ for all $1 \leq i \leq n$, we have $I(S_i) \in M$ which implies $ \bigast_{i=1}^n I(S_i) \in M$, by the closure of $M$ under the fusion. Finally, since $M$ is closed under modus ponens, $I(S) \in M$ which implies $G \vdash S$, by Definition \ref{CalculusForLogic}.
\end{proof}

\begin{cor}\label{MainCor}
Let $L$ be a logic over $\mathcal{L} \in \{\mathcal{L}_I^b, \mathcal{L}_I^u\}$ and $G$ a single-conclusion (resp. multi-conclusion) semi-analytic calculus for $L$. Then, 
\begin{itemize}
\item 
if $\mathcal{L} =\mathcal{L}_I^{b}$ and $\mathsf{IMALL} \subseteq L$ (resp. $\mathsf{MALL} \subseteq L$),  then $L$ has the CIP, and
\item 
if $\mathcal{L} =\mathcal{L}_I^{u}$ and $\mathsf{FL_e} \subseteq L$ (resp. $\mathsf{CFL_e} \subseteq L$),  then $L$ has the CIP.
\end{itemize}
\end{cor}
\begin{proof}
By Lemma \ref{FLAdmissibility}, it is clear that $G$ is strong. Therefore, by Theorem \ref{MainThm}, $G$ has the single-conclusion (resp. multi-conclusion) interpolation. Thus, by Theorem \ref{MaeharaImpliesInterpolation}, the logic $L$ enjoys the CIP.
\end{proof}

\section{Applications}
In this section, we address two types of applications of Corollary \ref{MainCor}. On the positive side, we provide a uniform and modular method to establish CIP for several multimodal substructural logics, only by checking the syntactical form of the axioms and rules of their sequent calculi.
On the negative side, we leverage the rarity of the CIP to offer a formal proof for the intuition that having a ``nice" sequent calculus is infrequent. For the latter, we employ our approximation of semi-analytic calculi as the formalization of niceness.

\begin{cor}\label{ApplicationInt1} 
The logic of $\mathbf{G}_{(I, \preccurlyeq, F)}$ or $\mathbf{CG}_{(I, \preccurlyeq, F)}$ over $\mathcal{L} \in\{\mathcal{L}_I^b, \mathcal{L}_I^{u}\}$, for any suitable SDML $(I, \preccurlyeq, F)$ enjoys the CIP.
Concrete examples are the logics in Example \ref{exm: logics} and the logics of sequent calculi in Theorem \ref{thm: cut mardom}.
\end{cor}
\begin{proof}
The cut-free sequent calculi for the logics are semi-analytic. Note that the rules $(\mathsf{T}_\alpha), (\mathsf{W}!^\alpha)$, and $(\mathsf{C}!^\alpha)$ are semi-analytic. Therefore, by Corollary \ref{MainCor}, we get the CIP for their logics. 
\end{proof}

The CIP for some of the logics in Corollary \ref{ApplicationInt1} are proved before. For instance, the CIP for $\mathsf{CLL}$ and $\mathsf{MALL}$ is proved in \cite{roorda} and for $\mathsf{FL_e}$ in \cite{OnoKomori}. However, to the best of our knowledge, the CIP for the logics of $\mathbf{G}_{(I, \preccurlyeq, F)}$ and $\mathbf{CG}_{(I, \preccurlyeq, F)}$ over $\mathcal{L} \in\{\mathcal{L}_I^b, \mathcal{L}_I^{u}\}$ in its full generality is new.

For negative applications, we first recall some families of substructural, intermediate and modal logics. We assume the basic algebraic notions such as lattices, monoids, and varieties are defined as usual (see \cite{ono,March}). 

\begin{dfn}
A \emph{pointed commutative residuated lattice} is a structure $\mathbb{A} = \langle A , \wedge , \vee , * , \to , 0, 1 \rangle$, where $\wedge , \vee , *, \to $ are binary operations, $0,1$ are constants, $\langle A, \wedge , \vee \rangle$ is a lattice with the partial order $\leq$, $\langle A, *, 1 \rangle$ is a commutative monoid, and $x * y \leq z$ if{f} $x \leq y \to z$, for any $x, y, z \in A$.
For a single pointed commutative residuated lattice $\mathbb{A}$ and a class of pointed commutative residuated lattices $\mathcal{K}$, denote the varieties generated by $\mathbb{A}$ and $\mathcal{K}$, by $\mathcal{V}(\mathbb{A})$ and $\mathcal{V}(\mathcal{K})$, respectively.
\end{dfn}

\begin{table}[t]
 \begin{center}
\scalebox{0.9}{{\begin{tabular}{ |c| c| }
\hline
Name  & Equational condition \\
\hline
 prelinearity (prl) & $1 \leq (x \to y) \vee (y \to x)$ \\
\hline
 distributivity (dis) & $x \wedge (y \vee z) = (x \wedge y) \vee (x \wedge z)$\\
\hline
 involutivity (inv) & $\neg \neg x = x$ \\
\hline
 integrality (int) & $x \leq 1$\\
\hline
 boundedness (bd) & $0 \leq x$ \\
\hline
 idempotence (id) & $x = x * x$\\
\hline
 fixed point negation (fp) & $0=1$ \\
\hline
 divisibility (div) & $x * (x \to y) = y * (y \to x)$\\
\hline
 cancellation (can) & $x \to (x * y)= y$ \\
\hline
 restricted cancellation (rcan) & $1= \neg x \vee ((x \to (x * y)) \to y)$ \\
\hline
 noncontradiction (nc) & $x \wedge \neg x \leq 0$\\
\hline
\end{tabular}}}
\caption{\small{Some equational conditions for pointed commutative residuated lattices.}}\label{table: equational conditions}
\end{center}
\end{table}

\begin{table}[t]
\begin{center}
\scalebox{0.9}{\begin{tabular}{ |c| c| }
\hline
Name of the logic & Axioms (including $(prl)$, $(dis)$) \\
\hline
 unbounded uninorm $(\mathsf{UL^{-}})$ & \\
\hline
 unbounded involutive uninorm  $(\mathsf{IUL^{-}})$  & $(inv)$ \\
\hline
  monoidal t-norm $(\mathsf{MTL})$ & $(int), (bd)$ \\
\hline
  strict monoidal t-norm  $(\mathsf{SMTL})$ & $(int), (bd), (nc)$ \\
\hline
  involutive monoidal t-norm  $(\mathsf{IMTL})$ & $(int), (bd), (inv)$ \\
\hline
  basic fuzzy  $(\mathsf{BL})$ & $(int), (bd), (div)$ \\
\hline
  G\"{o}del  $(\mathsf{G})$ & $(int), (bd), (id)$ \\
\hline
  \L ukasiewicz  (\textsf{\L}) & $(int), (bd), (div), (inv)$ \\
\hline
  product  $(\mathsf{P})$ & $(int), (bd), (div), (rcan)$ \\
\hline
  cancellative hoop $(\mathsf{CHL})$ & $(int), (fp), (div), (can)$ \\
\hline
  unbounded uninorm mingle  $(\mathsf{UML^{-}})$ & $(id)$ \\
\hline
  $R$-mingle with unit $(\mathsf{RM^e})$ & $(id), (inv)$ \\
\hline
  unbounded involutive uninorm mingle  $(\mathsf{IUML^{-}})$ & $(id), (inv), (fp)$ \\
\hline
  abelian $(\mathsf{A})$ & $(inv), (fp), (can)$\\
\hline
\end{tabular}}
\caption{\small{In each logic, the axioms $(prl)$ and $(dis)$ are present.}}\label{table: logics}
\end{center}
\end{table}

\noindent Tables \ref{table: equational conditions} and \ref{table: logics} (adapted from \cite{March}) present equational conditions to define varieties of pointed commutative residuated lattices and their logics in the usual sense over $\mathcal{L}^u_{\varnothing}$. Also, for any $n > 1$, let ${L}_n = \{0, \frac{1}{n-1}, \dots , \frac{n - 2}{n - 1}, 1 \}$ and
\begin{center}
$\mathbb{L}_n = \langle L_n , min, max , *_{\text{\L}}, \to_{\text{\L}}, 1, 0 \rangle
 \quad , \quad
\mathbb{G}_n = \langle L_n , min, max , min, \to_{G}, 1, 0 \rangle$    
\end{center}
where $x *_{\text{\L}} y = max \{0, x +y-1\}$, $x \to_{\text{\L}} y = min \{1, 1-x+y\}$, and $x \to_G y$ is $y$ if $x > y$, otherwise 1. Define \textsf{\L}$_n$ and $\mathsf{G_n}$ for $n >1$ as the logics of $\mathcal{V}(\mathbb{L}_n)$ and $\mathcal{V}(\mathbb{G}_n)$, respectively.
$\mathsf{R}$ is the logic of the variety of distributive pointed commutative residuated lattices satisfying $x \leq x*x$ and $\neg \neg x=x$ for any $x$. \\
Finally, on the set of integers $\mathbb{Z}$, define
\begin{center}
$  x* y =
\begin{cases}
x \wedge y & \text{if} \; |x|=|y| \\ 
y & \text{if} \; |x|<|y| \\ 
x & \text{if} \; |y|<|x|
\end{cases} 
\hspace{25pt}
x \to y =
\begin{cases} 
(-x) \vee y & \text{if} \; x \leq y \\ 
(-x) \wedge y & \text{otherwise}
\end{cases}  $
\end{center}
where $\wedge$ and $\vee$ are min and max, respectively, and $|x|$ is the absolute value of $x$. Define $\mathsf{RM^e_n}$ as the logic of $\mathcal{V}(\mathbb{S}_n)$, where

\begin{flushleft}
$ \mathbb{S}_{2m}= \langle \{-m, -m+1, \dots, -1, 1, \dots, m-1, m\}, \wedge, \vee, *, \to, 1, -1 \rangle \;\; (m \geq 1) $,

$ \mathbb{S}_{2m+1}= \langle \{-m, -m+1, \dots, -1, 0, 1, \dots, m-1, m\}, \wedge, \vee, *, \to, 0, 0 \rangle \;\; (m \geq 0)$.
 \end{flushleft}

\noindent Recall the intermediate logics $\mathsf{KC}=\mathsf{IPC}+\neg p \vee \neg \neg p$, $\mathsf{Bd_2}=\mathsf{IPC}+ p \vee (p \to (q \vee \neg q))$, $\mathsf{Sm}=\mathsf{Bd_2}+\neg p \vee \neg \neg p$, and $\mathsf{GSc}= \mathsf{Bd_2}+ (p \to q) \vee (q \to p) \vee (p \leftrightarrow \neg q)$. Finally, recall that $\mathsf{S4}$ is the logic of the system $\mathbf{CLL}+\{Lc, Rc, Lw, Rw\}$ and $\mathsf{Grz}=\mathsf{S4}+ \bigl(!^{\alpha} (!^{\alpha} (p \to !^{\alpha} p)\to p ) \to p \bigr)$ over the language $\mathcal{L}^u_{\{\alpha\}}$. 
 
For negative applications, we exploit the failure of the CIP for several logics extending $\mathsf{FL_e}$ (resp. $\mathsf{CFL_e}$) as shown for example in \cite{Ghil1,Max,March,Urq}.
By Corollary \ref{MainCor} these logics do not have a single-conclusion (resp. both single-conclusion and multi-conclusion) semi-analytic calculus. The exact statement is the subsequent corollary, including citations to the pertinent papers establishing the lack of the CIP for the logics under consideration.

\begin{cor} \label{Cor_Negative_Applications}
The following families of logics lack the CIP and hence cannot have a single-conclusion semi-analytic sequent calculus. The first family: 
\item[$\bullet$]
\cite{ba,luk,Urq}\cite[Corollary 3.2]{March} $\mathsf{UL^-}$, $\mathsf{MTL}$, $\mathsf{SMTL}$, $\mathsf{BL}$, $\mathsf{P}$, $\mathsf{CHL}$, $\mathsf{G}_n$ (for $n \geq 4$);
\item[$\bullet$]
\cite{Max}
none of the consistent extensions of $\mathsf{IPC}$, except for possibly $\mathsf{IPC}$, $\mathsf{G}$, $\mathsf{KC}$, $\mathsf{Bd_2}$, $\mathsf{Sm}$, $\mathsf{GSc}$ and $\mathsf{CPC}$; 
\item[$\bullet$]
\cite{ba,mo}\cite[Theorem 3.3]{March}
none of the consistent $\mathsf{BL}$-extensions, except for possibly $\mathsf{G}$, $\mathsf{G3}$ and $\mathsf{CPC}$;

and the second family:

\item[$\bullet$]
\cite{ba,luk,Urq}\cite[Corollary 3.2]{March}
$\mathsf{R}$, $\mathsf{A}$, $\mathsf{IUL^-}$, $\mathsf{IMTL}$, \textsf{\L}, \textsf{\L}$_n$ (for $n \geq 3$);
\item[$\bullet$]
\cite{max91} none of the consistent extensions of $\mathsf{S4}$, except for at most 37 of them;
\item[$\bullet$]
\cite{max91}
none of the consistent extensions of $\mathsf{Grz}$, except for at most 6 of them;
\item[$\bullet$]
\cite[Corollary 3.5]{March}
none of the consistent $\mathsf{IMTL}$-extensions, and consequently, none of the consistent extensions of
 $\textsf{\L}$, except for $\mathsf{CPC}$ \cite{komori81};
\item[$\bullet$]
\cite[Theorem 4.9]{March} none of the consistent extensions of $\mathsf{RM^e}$, except for possibly $\mathsf{RM^e}$, $\mathsf{IUML^-}$, $\mathsf{CPC}$, $\mathsf{RM^e_3}$, $\mathsf{RM^e_4}$, $\mathsf{CPC} \cap \mathsf{IUML^-}$, $\mathsf{RM^e_4} \cap \mathsf{IUML^-}$, and $\mathsf{CPC} \cap \mathsf{RM^e_3}$. This includes:
\begin{itemize}
\item[$(i)$]
$\mathsf{RM^e_n}$ for $n \geq 5$,
\item[$(ii)$]
$\mathsf{RM^e_{2m}} \cap \mathsf{RM^e_{2n+1}}$ for $n \geq m \geq 1$ with $n \geq 2$.,
\item[$(iii)$]
$\mathsf{RM^e_{2m}} \cap \mathsf{IUML^-}$ for $m \geq 3$;
\end{itemize}
\item[$\bullet$]
\cite[Corollary 3.14]{fussner} continuum-many
extensions of $\mathsf{MALL}$ and $\mathsf{CLL}$.

The logics in the second family do not have a multi-conclusion semi-analytic calculus either.
\end{cor}

\section{Conclusion and future work}
We presented a general form of rules and axioms, \emph{semi-analytic} and \emph{focused}, respectively, to formalize the nice axioms and rules common in the proof-theoretic literature. We showed that any strong enough calculus only consisting of these rules and axioms and possibly some basic modal rules enjoys the CIP. This provides a uniform framework to prove the CIP for many logics by looking into the pure form of the axioms and rules in their sequent calculi. More interestingly though, as the CIP is a rare property, our result indicates that many logics cannot have such a nice calculus. This provides a formal proof for the well-known intuition in the proof-theoretic community that nice sequent calculi are rare to find.

The logical property we considered in this paper is the Craig interpolation property. There are other invariants one can use for the study of the existence problem (as discussed in Introduction). Some related works in this area are \cite{khod1,khod2,khod3,tabrah1,tabrah2}, considering the uniform interpolation property and admissibility of certain rules, respectively. As for future work, it would be desirable to extend the method to sequent calculi that have non-normal modal rules or allow analytic applications of the cut rule. It would also be interesting to investigate general forms of rules in other calculi than sequent calculi, for instance, hypersequent and display calculi \cite{avron1996,Belnap,brother,AgataWansing,Gore2}, nested calculi \cite{bru,bru2}, labelled calculi \cite{Negri,vigano}, bunched sequent calculi \cite{Dunn,mints}, bunched hypersequent calculi \cite{AgataRevantha}, and
multi-conclusion nested sequent calculi \cite{kuz}. 
Finally, we would like to extend the method to first-order theories and investigate other logical properties such as Skolemization.\\

\noindent \textbf{Acknowledgements.} 
We are grateful to Rosalie Iemhoff for bringing this 
line of research to our attention, for her generosity in sharing her ideas on
the subject of  universal proof theory and for enlightening discussions. We appreciate valuable comments and suggestions by Hiroakira Ono, Pavel Pudl\'{a}k, Dale Miller, Elaine Pimentel, George Metcalfe, Revantha Ramanayake, and Stepan Kuznetsov. 

\bibliographystyle{plain}
\bibliography{main}

\end{document}